\documentclass[reprint,notitlepage,amsmath,amssymb,aps,pra,onecolumn,floatfix]{revtex4-1}
\usepackage[margin=1.0in]{geometry}
\usepackage{epsfig,ulem,amssymb,soul}
\usepackage{hyperref}
\usepackage{graphicx}
\usepackage{epstopdf}
\usepackage{amsfonts,amsmath}
\usepackage{bm}
\usepackage{setspace}
\usepackage{comment}
\usepackage{siunitx}
\usepackage{mathrsfs}
\usepackage{multirow}
\usepackage{subcaption}
\usepackage[hang,flushmargin,bottom]{footmisc}
\usepackage{mathtools}
\usepackage{lipsum}
\usepackage{array}
\usepackage{enumerate}
\usepackage{ifthen}
\usepackage{leftidx}
\usepackage{relsize}
\usepackage{braket}
\usepackage{float}
\DeclarePairedDelimiter{\ceil}{\lceil}{\rceil}
\usepackage[autostyle]{csquotes}
\usepackage[dvipsnames]{xcolor}

\newcommand{\etal}{\textit{et al}.~}
\providecommand{\keywords}[1]{\textbf{\textit{Keywords-}} #1}

\begin{document}
\title{Fractional-Order Models for the Static and Dynamic Analysis of Nonlocal Plates}
\author{Sansit Patnaik$^\dagger$}
\author{Sai Sidhardh}
\author{Fabio Semperlotti$^\dagger$}
\affiliation{School of Mechanical Engineering, Ray W. Herrick Laboratories, Purdue University, West Lafayette, IN 47907}

\begin{abstract}
This study presents the analytical formulation and the finite element solution of fractional order nonlocal plates under both Mindlin and Kirchoff formulations. By employing consistent definitions for fractional-order kinematic relations, the governing equations and the associated boundary conditions are derived based on variational principles. Remarkably, the fractional-order nonlocal model gives rise to a self-adjoint and positive-definite system that accepts a unique solution.
Further, owing to the difficulty in obtaining analytical solutions to this fractional-order differ-integral problem, a 2D finite element model for the fractional-order governing equations is presented. Following a thorough validation with benchmark problems, the 2D fractional finite element model is used to study the static as well as the free dynamic response of fractional-order plates subject to various loading and boundary conditions. It is established that the fractional-order nonlocality leads to a reduction in the stiffness of the plate structure thereby increasing the displacements and reducing the natural frequency of vibration of the plates. Further, it is seen that the effect of nonlocality is stronger on the higher modes of vibration when compared to the fundamental mode. These effects of the fractional-order nonlocality are noted irrespective of the nature of the boundary conditions. More specifically, the fractional-order model of nonlocal plates is free from boundary effects that lead to paradoxical predictions such as hardening and absence of nonlocal effects in classical integral approaches to nonlocal elasticity. This consistency in the predictions is a result of the well-posed nature of the fractional-order governing equations that accept a unique solution.\\

\noindent\keywords{Fractional Calculus, Nonlocal Plates, Variational Calculus, Finite Element Method}
\noindent$^\dagger$ All correspondence should be addressed to: \textit{spatnai@purdue.edu} or \textit{fsemperl@purdue.edu}
\end{abstract}
\maketitle
\section{Introduction}
\label{sec: Introduction}
In the recent years, following the rapid growth in many engineering fields including, but not limited to, functionally graded materials (FGM), composites, nanotechnology, and MEMS, the modeling of the static as well as the dynamic response of complex slender structures has received considerable attention. FGMs \cite{udupa2014functionally,naebe2016functionally} and sandwiched designs have found many useful applications in the design of macroscale structures such as those involved in naval and automotive systems, as well as lightweight structures, such as those employed in space and aeronautic applications. Similarly, micro- and nano-structures such as thin films, carbon nanotubes, monolayer graphene sheets, and micro tubules have demonstrated far-reaching applications in atomic devices, micro/nano-electromechanical devices, sensors, and even biological implants \cite{wang2011mechanisms,pradhan2009small,narendar2011buckling}. Many of these complex structures often employ a combination of slender and thin-walled structures like beams, plates, and shells, whose response was often shown to be significantly impacted by size-dependent effects, also referred to as nonlocal effects. In the case of macrostructures, these nonlocal effects were shown to be originated from material heterogeneities, interactions between layers (e.g. FGMs or composites) or unit cells (e.g. periodic media) \cite{romanoff2016using,tarasov2013review,hollkamp2019analysis,patnaik2019generalized}. In other terms, nonlocal governing equations for macrostructures often result from a process of homogenization of the initial inhomogeneous system. In the case of nano- and micro-structures, these size-dependent effects have been traced back to the existence of surface and interface stresses due to either nonlocal atomic or Van der Waals interactions \cite{wang2011mechanisms}. Further, geometric effects such as, for example, changes in curvature have also been shown to induce nonlocal size-dependent effects \cite{sudak2003column,pradhan2009small,sahmani2018nonlocal}.
It appears that the ability to accurately model these nonlocal effects as part of the structural response is paramount in many engineering applications. In order to capture scale effects, nonlocal continuum theories were developed. 

From a general perspective, nonlocal continuum theories enrich the classical (local) governing equations with information of the behavior of points within a prescribed distance; the latter typically indicated as the horizon of influence or horizon of nonlocality. The key principle behind nonlocal theories relies on the idea that all the particles located inside the horizon influence one another by means of long range cohesive forces. Seminal works from Kro\"ner \cite{kroner1967elasticity}, Eringen \cite{eringen1972nonlocal}, and several other authors \cite{nowinski1984nonlocal,lim2015higher} have explored the role of nonlocality in elasticity and laid its theoretical foundation. 
The mathematical description of nonlocal continuum theories relies on the introduction of additional contributions in terms of gradient or integrals of the strain field in the constitutive equations. This leads to so-called “weak” gradient methods or “strong” integral methods, respectively. 
Gradient elasticity theories \cite{peerlings2001critical,sidhardh2018exact,sidhardh2018element} account for the nonlocal behavior by introducing strain gradient dependent terms in the stress-strain constitutive law. Integral methods \cite{polizzotto2001nonlocal,bavzant2002nonlocal} model nonlocal effects by defining the constitutive law in the form of a convolution integral between the strain and the spatially dependent elastic properties over the horizon of nonlocality. 

Several researchers have used the above mentioned nonlocal theories to model the response of nonlocal beams and plates with particular attention to their applications in micro- and nano-devices \cite{lu2007non,reddy2010nonlocal} where nonlocal effects are typically more noticeable. In particular, the effect of the nonlocality on the buckling load of the plates \cite{pradhan2009small,narendar2011buckling,sahmani2018unified} as well as the vibration response of the plates \cite{barati2018nonlinear} have been extensively studied. 
In this context, Challamel \etal \cite{challamel2016buckling,zhang2014eringen} have developed a phenomenological nonlocal model from a lattice-based nonlocal model from analytic continuation of the plate lattice equations and have used it to analyze the effect of nonlocality on the buckling loads of a nonlocal plate. 
All these studies have shown that the introduction of nonlocality leads to a decrease in the stiffness of the structure which translates to higher static displacements \cite{yan2015exact,sahmani2018nonlocal}, lower buckling loads, and lower frequencies of vibration of the nonlocal structures. We merely note that a mixture of analytical \cite{duan2007exact,yan2015exact} as well numerical methods have been employed to determine the response of the nonlocal plates in the aforementioned studies. The numerical strategies included perturbation methods \cite{barati2018nonlinear,sahmani2018nonlocal}, differential quadrature methods \cite{pradhan2009small,golmakani2014nonlinear}, Galerkin methods \cite{phadikar2010variational,anjomshoa2013application} as well as spectral collocation methods \cite{ma2016vibration}.

Although these classical studies on nonlocal elasticity have been able to address several features of the response of size-dependent nonlocal structures, they encounter some key shortcomings. 
Gradient theories provide a satisfactory description of the material micro structure, but they introduce serious difficulties when enforcing the boundary conditions associated with the strain gradient-dependent terms \cite{peerlings2001critical,aifantis2003update}. 
On the other side, the integral methods are better suited to deal with boundary conditions but require the attenuation functions to have a positive Fourier transform everywhere in order to avoid instabilities \cite{bavzant1984instability,bavzant2002nonlocal}. 
Additionally, in both these classes of methods, the stress at any point cannot be obtained unless the strains in the neighbourhood of the particular point are known. In other terms, there exists no explicit relation for obtaining the stress at a given point from the strain at that particular point. This prevents the application of variational principles in these theories. More specifically, the basis of variational formulation is the principle of minimum total potential energy which is valid under the assumption that the stress at a point can be uniquely defined in terms of the strain at that point. Since an explicit relation between stress and strain components at a reference point cannot be found in the classical nonlocal theories, the principle of minimum potential energy cannot be applied to these theories. This aspect prevents the development of variational finite element methods to obtain numerical solutions to the classical nonlocal formulations. We note that the development of numerical methods is essential in this class of problems because the complex nonlocal governing equations associated with both gradient and integral formulations do not generally admit closed-form analytical solutions. Several researchers have developed inverse approaches in order to define a quadratic form of the total potential energy and then use it to obtain the response of the nonlocal structures via Galerkin or Ritz approximations \cite{phadikar2010variational,anjomshoa2013application}. In addition to the above shortcomings, classical integral approaches lead to mathematically ill-posed governing equations which leads to erroneous predictions such as the absence of nonlocal effects and the occurrence of hardening behavior for certain combinations of boundary conditions \cite{romano2017constitutive,romano2018nonlocal}. 
In this class of problems, the ill-posedness stems from the fact that the constitutive relation between the bending field and the curvature is a Fredholm integral of the first kind, whose solution does not generally exists and, if it exists, it is not necessarily unique \cite{romano2017constitutive,romano2018nonlocal}.

In recent years, fractional calculus has emerged as a powerful mathematical tool to model a variety of nonlocal and multiscale phenomena. Fractional derivatives, which are a differ-integral class of operators, are intrinsically multiscale and provide a natural way to account for nonlocal effects. In fact, the order of the fractional derivative dictates the shape of the power-law influence function (kernel of the fractional derivative) while its interval defines the horizon of its influence, i.e., the distance beyond which information is no longer accounted for in the derivative. 
As a result, time-fractional operators enable memory effects (i.e. the response of a system is a function of its past history) while space-fractional operators can account for nonlocal and scale effects. 
Given the multiscale nature of fractional operators, fractional calculus has found wide-spread applications in nonlocal elasticity. Riesz-type fractional derivatives have been shown to emerge as the continuum limit of discrete systems (e.g. such as chains and lattices) with power-law long-range interactions \cite{tarasov2013review}. Space-fractional derivatives have been used to formulate nonlocal constitutive laws \cite{drapaca2012fractional,carpinteri2014nonlocal,sumelka2014fractional,sumelka2015non,hollkamp2020application} as well as to account for microscopic interaction forces \cite{cottone2009elastic,di2008long}. Space-fractional derivatives have also been employed to capture attenuation including a variety of conditions such as interatomic nonlocal forces \cite{cottone2009elastic}, nonlocal stress-strain constitutive relations \cite{atanackovic2009generalized}, and even bandgaps in periodic media \cite{hollkamp2019analysis}.
Previous works conducted on the development of nonlocal continuum theories based on fractional calculus have highlighted that the differ-integral nature of the fractional operators allows them to combine the strengths of both gradient and integral based methods while at the same time addressing a few important shortcomings of the integer order formulations \cite{drapaca2012fractional,carpinteri2014nonlocal,sumelka2014fractional,patnaik2019generalized,patnaik2019FEM,sidhardh2020geometrically}. 

In this study, we build upon the fractional-order nonlocal continuum model proposed in \cite{patnaik2019generalized} to develop a fractional-order model for nonlocal plates. 
The overall goal of this study is three fold. 
First, we derive the governing equations for the nonlocal plates in a strong form by using variational principles. We will show that the fractional-order formulation allows the application of variational principles because the stress in the fractional-order formulation can be uniquely and explicitly determined from the strain at the particular point. In fact, nonlocality will be introduced into the plate through nonlocal fractional-order kinematics and not through the classical integral constitutive relations between the stress and the strain.
Second, we formulate a fully consistent and highly accurate 2D fractional-order finite element method (f-FEM) by extending the work in \cite{patnaik2019FEM} on 1D fractional-order beams, to numerically investigate the response of the fractional-order nonlocal plates. Although several FE formulations for fractional-order equations have been proposed in the literature, they are based on Galerkin or Petrov-Galerkin methods that are capable of solving 1D hyperbolic and parabolic differential equations involving transport processes \cite{jin2016petrov,li2017galerkin}. 
We develop a Ritz FEM that is capable of obtaining the numerical solution of the fractional-order plate governing equations. We highlight here that although Ritz FEMs for classical nonlocal elasticity problems have been developed in the literature, they do not extend to fractional-order nonlocal modeling, because the attenuation function capturing the nonlocal interactions in the fractional-order model involves a singularity within the kernel \cite{podlubny1998fractional}. Moreover, the necessity and flexibility of the f-FEM developed here becomes clear from the complexities involved in the handling of integral boundary conditions \cite{fernandez2016bending}.
Finally, we use the developed 2D f-FEM to analyze the effect of nonlocality on the static and free vibration response of the fractional-order plates. We will show that independently from the boundary conditions, the fractional-order theory predicts a consistent softening behaviour for the fractional-order plates as the nonlocality degree increases. This latter aspect is unlike the paradoxical predictions of hardening or of the absence of nonlocal effects predicted by classical integral nonlocal approaches \cite{challamel2014nonconservativeness,khodabakhshi2015unified,romano2017constitutive,romano2018nonlocal,barretta2019stress} for certain combinations of boundary conditions.

The remainder of the paper is structured as follows: first, we introduce the fractional-order formulation used in this study to model nonlocal elasticity. Next, we derive the governing equations of fractional-order Mindlin as well as fractional-order Kirchoff plates in strong form using variational principles. Further, we derive a strategy for obtaining the numerical solution to the plate governing equations using 2D f-FEM. Finally we validate the 2D f-FEM, establish its convergence, and then use it to analyze the effect of the fractional-order nonlocality on the static and free vibration response of plates under different types of loading conditions.

\section{Nonlocal Elasticity via Fractional Calculus}
\label{sec: Nonlocal Elasticity via Fractional Calculus}
Previous works conducted on the development of nonlocal continuum theories based on fractional calculus have highlighted its ability to combine the strengths of both gradient and integral based methods while simultaneously addressing a few important shortcomings of these integer-order formulations \cite{drapaca2012fractional,carpinteri2014nonlocal,sumelka2014fractional,patnaik2019generalized,patnaik2019FEM,sidhardh2020geometrically}. Recall that the key shortcomings included the requirement of higher-order essential boundary conditions in gradient based methods, the need for a kernel that is always positive in integral methods, and the inability of both methods to leverage variational principles. To this regard, note that the kernel used in fractional derivatives is positive everywhere \cite{podlubny1998fractional}. Unlike gradient elasticity methods, additional essential boundary conditions are not required when using Caputo fractional derivatives \cite{hollkamp2019analysis,patnaik2019generalized}. Further, we will show how variational principles can be used in the fractional-order formulation of nonlocality by using a specific strategy that involves defining a fractional-order (nonlocal) deformation gradient tensor.
The nonlocal plate theory presented in this work builds on the fractional-order nonlocal continuum formulation presented in \cite{patnaik2019generalized}. In the following, we briefly review the fractional-order nonlocal continuum model presented in \cite{patnaik2019generalized} which provides the foundation to develop the fractional-order nonlocal plate theory.

\begin{figure}[h]
	\centering
	\includegraphics[scale=0.45]{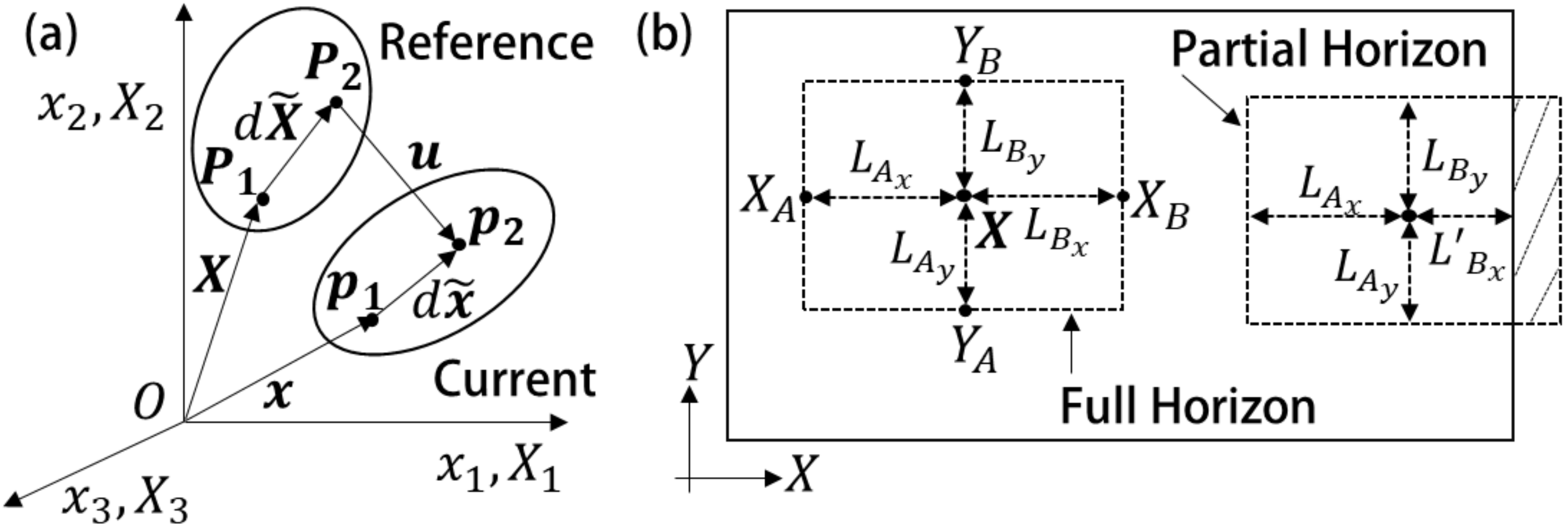}
	\caption{\label{fig1} 
	(a) Schematic indicating the infinitesimal material and spatial line elements in the nonlocal medium under the displacement field $\textbf{u}$. (b) Horizon of nonlocality and length scales at different material points. The nonlocal model can account for a partial (i.e. asymmetric) horizon condition that occurs for points ${\textbf{X}}$ close to a boundary or an interface. The schematic is adapted from \cite{patnaik2019generalized}.}
\end{figure}

In analogy with the traditional approach to continuum mechanics, we perform the deformation analysis of a nonlocal solid by introducing two configurations, namely, the reference (undeformed) and the current (deformed) configurations. The motion of the body from the reference configuration (denoted as $\textbf{X}$) to the current configuration (denoted as $\textbf{x}$) is assumed as:
\begin{equation}
\label{eq: motion_description}
\textbf{x}=\bm{\Psi}(\textbf{X},t)
\end{equation}
such that $\bm{\Psi}(\textbf{X},t)$ is a bijective mapping operation. The relative position of two point particles located at $P_1$ and $P_2$ in the reference configuration of the nonlocal medium is denoted by ${\mathrm{d}\tilde{\textbf{X}}}$ (see Fig.~(\ref{fig1})). After deformation due to motion $\bm{\Psi}(\textbf{X},t)$, the particles occupy the new positions $p_1$ and $p_2$, such that the relative position vector between them is ${\mathrm{d}\tilde{\textbf{x}}}$. It appears that $\mathrm{d}\tilde{\textbf{X}}$ and $\mathrm{d}\tilde{\textbf{x}}$ are the material and spatial differential line elements in the nonlocal medium, conceptually analogous to the classical differential line elements $\mathrm{d}{\textbf{X}}$ and $\mathrm{d}{\textbf{x}}$ in a local medium representation.

The mapping operation described in Eq.~(\ref{eq: motion_description}) captures the nonlocal behavior of the solid by leveraging fractional calculus formulation.
More specifically, the differential line elements of the nonlocal medium are modeled by imposing a fractional-order transformation on the classical differential line elements as follows:
\begin{subequations}
\label{eq: Fractional_F}
\begin{equation}
\mathrm{d}\tilde{\textbf{x}}=\big[D^{\alpha}_\textbf{X}\bm{\Psi}(\textbf{X},t)\big]\mathrm{d} \textbf{X}=\big[\tilde{\textbf{F}}_{X}(\textbf{X},t)\big]\mathrm{d}\textbf{X}
\end{equation}
\begin{equation}
\mathrm{d}\tilde{\textbf{X}}=\big[D^{\alpha}_\textbf{x}\bm{\Psi}^{-1}(\textbf{x},t)\big]\mathrm{d} \textbf{x}=\big[\tilde{\textbf{F}}_{x}(\textbf{x},t)\big]\mathrm{d}\textbf{x}
\end{equation}	
\end{subequations}
where $D^{\alpha}_\square\square$ is a space-fractional derivative whose details will be presented below.
Given the differ-integral nature of the space-fractional derivative, the differential line elements $\mathrm{d}\tilde{\textbf{X}}$ and $\mathrm{d}\tilde{\textbf{x}}$ have a nonlocal character.
Using the definitions for $\mathrm{d}\tilde{\textbf{X}}$ and $\mathrm{d}\tilde{\textbf{x}}$, the fractional deformation gradient tensor $\overset{\alpha}{\textbf{F}}$ with respect to the nonlocal coordinates is obtained in \cite{patnaik2019generalized} as:
\begin{equation}
\label{eq: Fractional_F_net}
\frac{\mathrm{d}\tilde{\textbf{x}}}{\mathrm{d}\tilde{\textbf{X}}}=\mathop{\textbf{F}}^{\alpha}=\tilde{\textbf{F}}_{X}\textbf{F}^{-1}\tilde{\textbf{F}}_{x}^{-1}
\end{equation}
where $\textbf{F}$ is the classical deformation gradient tensor given as $\textbf{F}=\mathrm{d}\textbf{x}/\mathrm{d}\textbf{X}$, in local and integer-order form.

The space-fractional derivative $D^{\alpha}_\textbf{X}\bm{\Psi}(\textbf{X},t)$ is taken according to a Riesz-Caputo (RC) definition with order $\alpha\in(0,1)$ defined on the interval $\textbf{X} \in (\textbf{X}_A,\textbf{X}_B) \subseteq \mathbb{R}^3 $ and given by:
\begin{subequations}
	\label{eq: RC_definition}
	\begin{equation}
	D^{\alpha}_\textbf{X}\bm{\Psi}(\textbf{X},t)=\frac{1}{2}\Gamma(2-\alpha)\big[\textbf{L}_{A}^{\alpha-1}~ {}^C_{\textbf{X}_{A}}D^{\alpha}_{\textbf{X}} \bm{\Psi}(\textbf{X},t) - \textbf{L}_{B}^{\alpha-1}~ {}^C_{\textbf{X}}D^{\alpha}_{\textbf{X}_{B}}\bm{\Psi}(\textbf{X},t)\big]
	\end{equation}
	\begin{equation}
	D^{\alpha}_{X_j}\Psi_i(\textbf{X},t)=\frac{1}{2}\Gamma(2-\alpha)\big[L_{A_j}^{\alpha-1}~{}^C_{X_{A_j}}D^{\alpha}_{X_j} \Psi_i(\textbf{X},t) - L_{B_j}^{\alpha-1}~{}^C_{X_j}D^{\alpha}_{X_{B_j}} {\Psi_i}(\textbf{X},t)\big]
	\end{equation}
\end{subequations}
where $\Gamma(\cdot)$ is the Gamma function, and ${}^C_{\textbf{X}_{A}}D^{\alpha}_{\textbf{X}}\bm{\Psi}$ and ${}^C_{\textbf{X}}D^{\alpha}_{\textbf{X}_{B}}\bm{\Psi}$ are the left- and right-handed Caputo derivatives of $\bm{\Psi}$, respectively. 
Before proceeding, we discuss certain implications of the above definition of the fractional-order derivative.
The interval of the fractional derivative $(\textbf{X}_A,\textbf{X}_B)$ defines the horizon of nonlocality (also called attenuation range in classical nonlocal elasticity) which is schematically shown in Fig.~(\ref{fig1}) for a generic point $\textbf{X}\in\mathbb{R}^2$. The length scale parameters $L_{A_j}^{\alpha-1}$ and $L_{B_j}^{\alpha-1}$ ensure the dimensional consistency of the deformation gradient tensor, and along with the term $\frac{1}{2}\Gamma(2-\alpha)$ ensure the frame invariance of the constitutive relations \cite{patnaik2019generalized}. As discussed in \cite{patnaik2019generalized}, the deformation gradient tensor introduced via Eq.~(\ref{eq: Fractional_F_net}) enables an efficient and accurate treatment of the frame invariance in presence of asymmetric horizons, material boundaries, and interfaces (see Fig.~(\ref{fig1})). 

In analogy with the classical strain measures, the nonlocal strain can be defined using the fractional-order differential line elements as $\mathrm{d}\tilde{\textbf{x}}d\tilde{\textbf{x}}-d\tilde{\textbf{X}}d\tilde{\textbf{X}}$. Using Eq.~(\ref{eq: Fractional_F_net}), the Lagrangian strain tensor in the nonlocal medium is obtained as:
\begin{equation}
\label{eq: strain_tensors}
	\mathop{\textbf{E}}^{\alpha}=\frac{1}{2}(\mathop{\textbf{F}}^{\alpha}{}^T\mathop{\textbf{F}}^{\alpha}-\textbf{I})
\end{equation}
where $\textbf{I}$ is the identity tensor. Using kinematic position-displacement relations, the expressions of the strains can be obtained in terms of the displacement gradients. The fractional displacement gradient is obtained using the definition of the fractional deformation gradient tensor from Eq.~(\ref{eq: motion_description}) and the displacement field $\textbf{U}(\textbf{X})=\textbf{x}(\textbf{X})-\textbf{X}$ as:
\begin{equation}
\label{eq: disp_grad_tesnor_Lag}
\nabla^\alpha {\textbf{U}}_X=\tilde{\textbf{F}}_{X}-\textbf{I}
\end{equation} 
The fractional gradient denoted by $\nabla^\alpha\textbf{U}_X$ is given as $ \nabla^\alpha\textbf{U}_{X_{ij}} = D^{\alpha}_{X_j}U_i$. Using the nonlocal strain defined in Eq.~(\ref{eq: strain_tensors}) and the fractional deformation gradient tensor $\overset{\alpha}{\textbf{F}}$ given in Eq.~(\ref{eq: Fractional_F_net}) together with Eq.~(\ref{eq: disp_grad_tesnor_Lag}), the relationship between the infinitesimal strain tensor and displacement gradient tensor is obtained as:
\begin{equation}
\label{eq: infinitesimal_fractional_strain}
{\bm{\epsilon}}=\frac{1}{2}\bigr(\nabla^\alpha {\textbf{U}}_X+\nabla^\alpha {\textbf{U}}_X^{T}\bigl)
\end{equation}
For small displacements, the above infinitesimal strain tensor has the same definition in both Eulerian and Lagrangian descriptions \cite{patnaik2019generalized}. Further, the stress tensor in the fractional-order model of the nonlocal medium has a form analogous to the local case as:
\begin{equation}
\label{eq: stress_equation}
\sigma_{ij} = C_{ijkl} \epsilon_{kl}
\end{equation}
where $C_{ijkl}$ denotes the constitutive matrix of the solid. We emphasize that the stress defined through the above equation is nonlocal in nature. This follows from the fractional-order definition of the deformation gradient tensor which is then reflected in the nonlocal strain as evident from Eq.~(\ref{eq: strain_tensors}). As expected, classical continuum mechanics relations are recovered when the order of the fractional derivative is set as $\alpha=1$.

Note that in the above presented fractional-order formulation, nonlocality has been modeled using fractional-order kinematic relations. More specifically, differential line elements in the undeformed and deformed nonlocal configurations were modeled using fractional-order deformation gradients which, in turn, were used to obtain the strain in the nonlocal medium. This definition of the strain has critical implications on the nonlocal formulation. Note that, given the strain at a point, the stress at the same point can be uniquely and explicitly determined by using Eq.~(\ref{eq: stress_equation}).
Recall that the basis of variational formulation is the principle of minimum total potential energy which is valid under the assumption that the stress at a point can be uniquely defined in terms of the strain at that point. It is immediate to deduce that the fractional-order model of nonlocality allows the application of variational principles. It is exactly this opportunity offered by the fractional operators that forms the basis for the development of a fractional-order model of nonlocal plates via the application of the extended Hamilton's principle. Additionally, this fractional-order formulation of nonlocality also ensures a quadratic form of the potential energy of the system as we will show in \S\ref{sec: Mindlin_Plates}. In the following, we will use this fractional-order formulation to model the response of both nonlocal Mindlin and Kirchoff plates. 

\section{Fractional-Order Modeling of Nonlocal Mindlin Plates}
\label{sec: Mindlin_Plates}
\begin{figure}[h]
	\centering
	\includegraphics[scale=0.5]{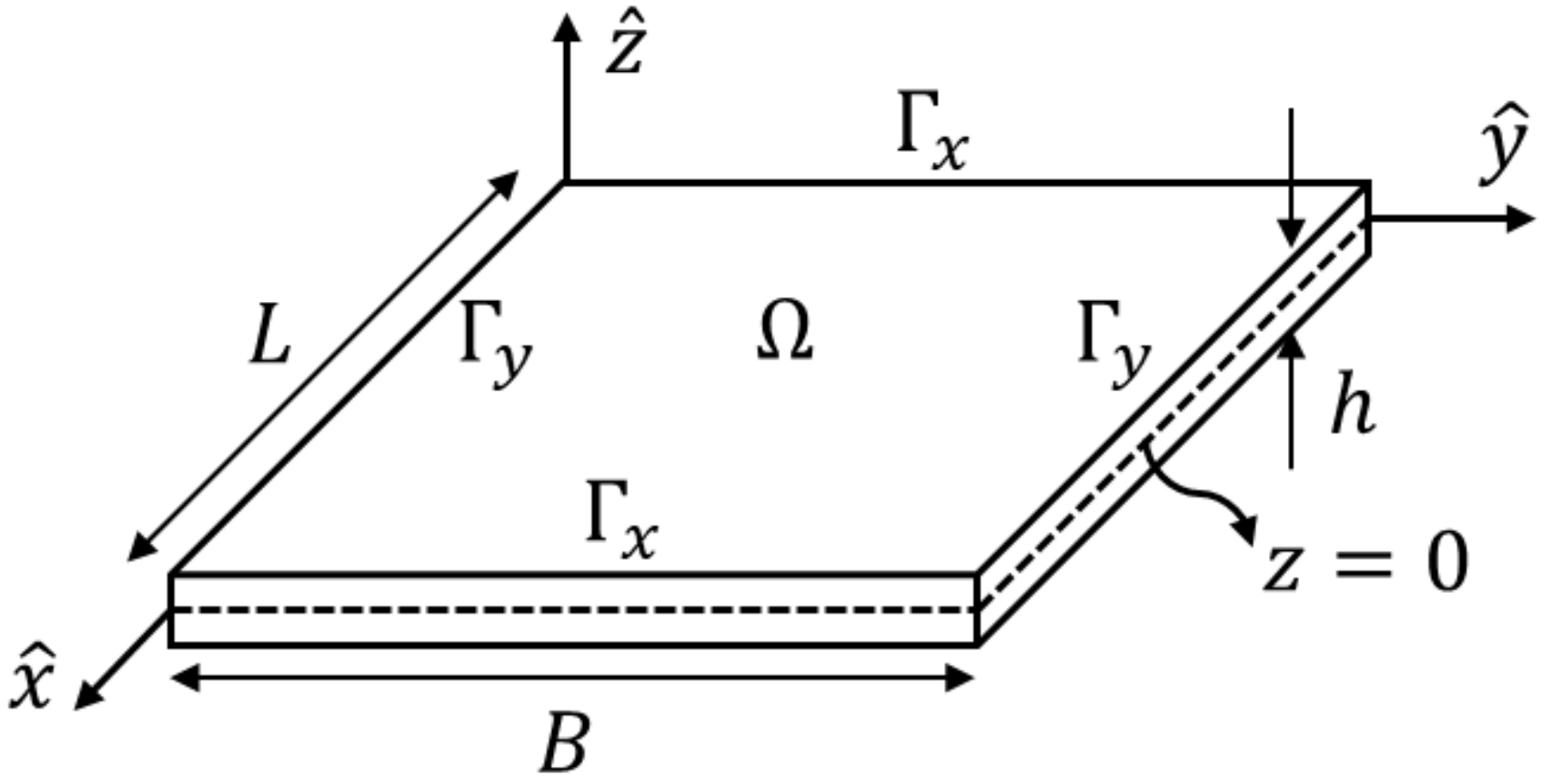}
	\caption{\label{fig: plate} Schematic of the rectangular plate showing key geometric parameters. The midplane of the plate is indicated by the $z=0$ plane of the domain $(\Omega)$. The boundaries of the domain are also indicated in the schematic.}
\end{figure}

We use the fractional-order continuum formulation presented above to develop a fractional-order analogue of the classical Mindlin plate formulation. A schematic of the undeformed rectangular plate along with the chosen Cartesian reference frame is illustrated in Fig.~(\ref{fig: plate}). The top surface of the plate is identified as $z=h/2$, while the bottom surface is identified as $z=-h/2$. The domain corresponding to the mid-plane of the plate (i.e., $z=0$) is denoted as $\Omega$, such that $\Omega=[0,L]\times[0,B]$ where $L$ and $B$ are the length and width of the plate, respectively. The domain of the plate is identified by the tensor product $\Omega\times[-h/2,h/2]$. The edges forming the boundary of the mid-plane of the plate are denoted as $\{\Gamma_x,\Gamma_y\}$ as shown in the Fig.~(\ref{fig: plate}).

For the chosen coordinate system, the in-plane and transverse components of the displacement field, denoted by $u(x,y,z,t)$, $v(x,y,z,t)$ and $w(x,y,z,t)$ at any spatial location $\textbf{x}(x,y,z)$, are related to the mid-plane displacements in the following manner:
\begin{subequations}
\label{eq: Mindlin_Kinematics}
    \begin{equation}
    u(x,y,z,t)=u_0(x,y,t) - z\theta_x(x,y,t)
    \end{equation}
    \begin{equation}
    v(x,y,z,t)=v_0(x,y,t) - z\theta_y(x,y,t)
    \end{equation}
    \begin{equation}
    w(x,y,z,t)=w_0(x,y,t)
    \end{equation}
\end{subequations}
where $u_0$, $v_0$, and $w_0$ are the mid-plane displacements of the plate along the $\hat{x}$, $\hat{y}$, and $\hat{z}$ directions. $\theta_x$ and $\theta_y$ are the rotations of the transverse normal about the $\hat{y}$ and $\hat{x}$ axes, respectively. In the interest of a more compact notation, the functional dependence of the displacement fields on the spatial and the temporal variables will be implied unless explicitly expressed to be constant. Based on the above displacement fields, the strain components in a fractional-order Mindlin plate are evaluated using Eq.~(\ref{eq: infinitesimal_fractional_strain}) as:
\begin{subequations}
\label{eq: Mindlin_relations}
\begin{equation}
    \epsilon_{xx}=D^\alpha_x u_0-zD^\alpha_x \theta_x
\end{equation}
\begin{equation}
    \epsilon_{yy}=D^\alpha_y v_0-zD^\alpha_y \theta_y
\end{equation}
\begin{equation}
    \gamma_{xy}=D^\alpha_y u_0+D^\alpha_x v_0-z(D^\alpha_y \theta_x+D^\alpha_x \theta_y)
\end{equation}
\begin{equation}
    \gamma_{xz}=D^\alpha_x w_0-\theta_x
\end{equation}
\begin{equation}
    \gamma_{yz}=D^\alpha_y w_0-\theta_y
\end{equation}
\begin{equation}
    \epsilon_{zz}=0
\end{equation}
\end{subequations}
Note that similarly to the classical Mindlin theory, the in-plane strain components $\{\epsilon_{xx},\epsilon_{yy},\gamma_{xy}\}$ vary in a linear fashion through the plate thickness, while the transverse shear strains $\{\gamma_{xz},\gamma_{yz}\}$ are constant through the thickness. The corresponding stresses are determined using the linear stress-strain relationships given in Eq.~(\ref{eq: stress_equation}). 

By using the above defined strain and stress fields, we derive the strong-form of the governing equations for the fractional-order plate using the extended Hamilton's principle:
\begin{equation}
    \label{eq: extended_hamiltons_principle}
    \int_{t_1}^{t_2}\left(\delta \mathcal{U} - \delta V - \delta T\right)\mathrm{d}t=0
\end{equation}
The nonlocal virtual strain energy $\delta \mathcal{U}$, the virtual work done by externally applied forces $\delta V$, and the virtual kinetic energy $\delta T$ are obtained as:
\begin{subequations}
\label{eq: Mindlin_virtual_quantities}
\begin{equation}
\label{eq: mindlin_virtual_strain_energy}
    \delta \mathcal{U} = \int_\Omega \bigg\{\int_{-\frac{h}{2}}^{\frac{h}{2}} \left[ \sigma_{xx}\delta\epsilon_{xx} + \sigma_{yy}\delta\epsilon_{yy} + \sigma_{xy}\delta\gamma_{xy} + \sigma_{xz}\delta\gamma_{xz} + \sigma_{yz}\delta\gamma_{yz} \right]\mathrm{d}z \bigg\} \mathrm{d} \Omega
\end{equation}
\begin{equation}
\label{eq: mindlin_virtual_work}
    \delta V = \int_\Omega \left[ F_x\delta u_0 + F_y \delta v_0 + F_z \delta w_0 + M_{\theta_x} \delta \theta_x + M_{\theta_y} \delta \theta_y \right]\mathrm{d} \Omega
\end{equation}
\begin{equation}
\label{eq: mindlin_virtual_kinetic_energy}
    \delta T = \int_\Omega \bigg\{\int_{-\frac{h}{2}}^{\frac{h}{2}} \rho \left[ \big(\dot{u}_0 - z\dot{\theta}_x \big)\big(\delta\dot{u}_0 - z\delta\dot{\theta}_x \big) + \big(\dot{u}_0 - z\dot{\theta}_x \big) \big(\delta\dot{u}_0 - z\delta\dot{\theta}_x \big) +\dot{w}_0 \delta \dot{w}_0 \right]\mathrm{d}z \bigg\} \mathrm{d} \Omega
\end{equation}
\end{subequations}
Note that $\mathrm{d}\Omega = \mathrm{d}x\mathrm{d}y$ for a rectangular plate. $\{F_x,F_y,F_z\}$ are the external loads applied in the $\hat{x}$, $\hat{y}$ and $\hat{z}$ directions, respectively, and $\{M_{\theta_x},M_{\theta_y}\}$ are the external moments applied about the $\hat{y}$ and $\hat{x}$ axes, respectively. By substituting the expression for the stress given in Eq.~(\ref{eq: stress_equation}) in Eq.~(\ref{eq: mindlin_virtual_strain_energy}), it is immediate that the potential energy of the fractional-order nonlocal plate is quadratic in nature. In fact, the use of the fractional-order formulation leads to a self-adjoint and positive definite system. The proof for the same can be found in \cite{patnaik2019FEM} where the fractional-order formulation introduced earlier in \S\ref{sec: Nonlocal Elasticity via Fractional Calculus} was used to model nonlocal beams. The same proof directly extends to the fractional-order plate formulations presented in the following and hence we do not provide it again here.

We first simplify the virtual strain energy in order to relieve the variations of any differentiation. By substituting the expressions for the strains in the expression for $\delta \mathcal{U}$ we obtain:
\begin{equation}
    \label{eq: Mindlin_virtual_strain_step1}
    \begin{split}
    \delta \mathcal{U} = \int_\Omega \Big[ N_{xx}\delta D^\alpha_x u_0 + M_{xx}\delta D^\alpha_x \theta_x + N_{yy}\delta D^\alpha_y v_0 + M_{yy}\delta D^\alpha_y \theta_y + N_{xy}\delta(D^\alpha_y u_0 + D^\alpha_x v_0) + \\ 
    M_{xy}\delta(D^\alpha_y \theta_x + D^\alpha_x \theta_y) + Q_{xz}\delta(D^\alpha_x w_0-\theta_x) + Q_{yz}\delta(D^\alpha_y w_0-\theta_y) \Big] \mathrm{d}\Omega
    \end{split}
\end{equation}
where the in-plane stress resultants $\{ N_{xx}, N_{yy}, N_{xy}\}$, the transverse stress resultants $\{Q_{xz}, Q_{yz}\}$ and the moment resultants $\{ M_{xx}, M_{yy}, M_{xy}\}$ are defined as:
\begin{subequations}
\label{eq: Mindlin_stress_resultants}
\begin{equation}
    \{ N_{xx}, N_{yy}, N_{xy}, Q_{xz}, Q_{yz}\} = \int_{-\frac{h}{2}}^{\frac{h}{2}} \{ \sigma_{xx}, \sigma_{yy}, \sigma_{xy}, K_s \sigma_{xz}, K_s\sigma_{yz}\} \mathrm{d}z
\end{equation}
\begin{equation}
    \{M_{xx}, M_{yy}, M_{xy}\} = \int_{-\frac{h}{2}}^{\frac{h}{2}} \{ -z\sigma_{xx}, -z\sigma_{yy}, -z\sigma_{xy} \}\mathrm{d}z
\end{equation}
\end{subequations}
where $K_s$ is the shear correction factor. The expression for $\delta \mathcal{U}$ in Eq.~(\ref{eq: Mindlin_virtual_strain_step1}) is further simplified using integration by parts and the definitions of the fractional derivatives. Here below, we provide the result of this simplification for only one term within the integral in Eq.~(\ref{eq: Mindlin_virtual_strain_step1}):
\begin{equation}
\label{eq: Mindlin_by_parts_1}
    \int_\Omega N_{xx} \left[ \delta D_{x}^{\alpha}u_0 \right] \mathrm{d}x \mathrm{d}y = -\int_\Omega \left[\mathfrak{D}_{x}^{\alpha} N_{xx}\right] \delta u_0 \mathrm{d}\Omega + \int_{\Gamma_x}\left[ I^{1-\alpha}_{x}N_{xx} \right] \delta u_0 \mathrm{d}y
\end{equation}
The detailed steps leading to the above simplification can be found in \cite{patnaik2019FEM} where a similar variational approach has been used in the context of 1D beams. In the above Eq.~(\ref{eq: Mindlin_by_parts_1}), $\mathfrak{D}_{x}^{\alpha}(\cdot)$ is the Riesz Riemann-Liouville derivative of order $\alpha$ which is defined as:
\begin{equation}
    \label{eq: r_rl_frac_der_def}
    \mathfrak{D}^{\alpha}_{x}\psi = \frac{1}{2}\Gamma(2-\alpha) \left[ l_{B_x}^{\alpha-1} \left({}^{RL}_{x-l_{B_x}} D^{\alpha}_{x} \psi\right) - l_{A_x}^{\alpha-1} \left( {}^{RL}_{x}D^{\alpha}_{x+l_{A_x}} \psi\right)\right]
\end{equation}
where $\psi$ is an arbitrary function and ${}^{RL}_{x-l_{B_x}}D^{\alpha}_{x}\psi$ and ${}^{RL}_{x}D^{\alpha}_{x+l_{A_x}}\psi$ are the left- and right-handed Riemann Liouville derivatives of $\psi$ to the order $\alpha$, respectively.
The Riesz fractional integral $I^{1-\alpha}_{x}(\cdot)$ is defined as:
\begin{equation}
\label{eq: reisz integral_def}
    I^{1-\alpha}_{x} \psi =\frac{1}{2}\Gamma(2-\alpha) \left[ l_{B_x}^{\alpha-1} {}_{x-l_{B_x}}I_{x}^{1-\alpha} \psi - l_{A_x}^{\alpha-1} {}_{x}I_{x+l_{A_x}}^{1-\alpha} \psi \right]
\end{equation}
where ${}_{x-l_{B_x}}I_{x}^{1-\alpha}\psi$ and ${}_{x}I_{x+l_{A_x}}^{1-\alpha}\psi$ are the left and right Riesz integrals to the order $\alpha$, respectively.
Note that the fractional derivative $\mathfrak{D}^{\alpha}_{x}(\cdot)$ and the fractional integral $I^{1-\alpha}_{x}(\cdot)$ are defined over the interval $(x-l_{B_x},x+l_{A_x})$ unlike the fractional derivative $D^{\alpha}_{x}(\cdot)$ which is defined over the interval $(x-l_{A_x},x+l_{B_x})$. This change in the terminals of the interval of the Riesz Riemann-Liouville fractional integral and derivative follows from the integration by parts technique used to simplify the variational integrals (see \cite{patnaik2019FEM}).
The remaining terms within the integral in Eq.~(\ref{eq: Mindlin_virtual_strain_step1}) can be simplified in a similar fashion. 

Note that the expression for the virtual work given in Eq.~(\ref{eq: mindlin_virtual_work}) can be directly used within Eq.~(\ref{eq: extended_hamiltons_principle}). Further, the expression of the virtual kinetic energy matches exactly its counterparts in the classical integer-order plate theory (see \cite{reddy2006theory}) and can be expressed as:
\begin{equation}
\label{eq: mindlin_virtual_kinetic_energy_simplified}
    \delta T = -\int_\Omega  \left[ I_0 \ddot{u}_0 \delta u_0 + I_0 \ddot{v}_0 \delta v_0 + I_0 \ddot{w}_0 \delta w_0 + I_2 \ddot{\theta}_x \delta \theta_x + I_2 \ddot{\theta}_y \delta \theta_y \right] \mathrm{d} \Omega
\end{equation}
where $I_0=\rho h$ and $I_2 = \rho h^3/12$. 

The expressions for the virtual quantities are substituted within the Hamilton's principle statement in Eq.~(\ref{eq: extended_hamiltons_principle}). The governing equations and the boundary conditions of the fractional-order Mindlin plate are then obtained by using the fundamental lemma of variational calculus as:
\begin{subequations}
\label{eq: Mindlin_GDE}
\begin{equation}
\mathfrak{D}^\alpha_x N_{xx} + \mathfrak{D}^\alpha_y N_{xy} + F_x = I_0\frac{\partial^2 u_0}{\partial t^2} 
\end{equation}
\begin{equation}
\mathfrak{D}^\alpha_x N_{xy} + \mathfrak{D}^\alpha_y N_{yy} + F_y = I_0\frac{\partial^2 v_0}{\partial t^2} 
\end{equation}
\begin{equation}
\mathfrak{D}^\alpha_x Q_{xz} + \mathfrak{D}^\alpha_y Q_{yz} + F_z = I_0\frac{\partial^2 w_0}{\partial t^2}
\end{equation}
\begin{equation}
\mathfrak{D}^\alpha_x M_{xx} + \mathfrak{D}^\alpha_y M_{xy} - Q_{xz} + M_{\theta_x} = I_2\frac{\partial^2\theta_x}{\partial t^2}
\end{equation}
\begin{equation}
\mathfrak{D}^\alpha_x M_{xy} + \mathfrak{D}^\alpha_y M_{yy} - Q_{yz} + M_{\theta_y} = I_2\frac{\partial^2\theta_y}{\partial t^2}
\end{equation}
\end{subequations}
The corresponding essential and natural boundary conditions are obtained as:
\begin{subequations}
\label{eq: Mindlin_BC}
\begin{equation}
    \delta u_0=0, ~\delta v_0=0, ~\delta w_0=0, ~\delta\theta_x=0, ~\delta \theta_y = 0 ~~ \forall~\Gamma_x \cup \Gamma_y
\end{equation}
\begin{equation}
    I_x^{1-\alpha}N_{xx}=0, ~I_x^{1-\alpha}N_{xy}=0, ~I_x^{1-\alpha}Q_{xz}=0, ~I_x^{1-\alpha}M_{xx}=0,
    ~I_x^{1-\alpha}M_{xy}=0 ~~ \forall~\Gamma_x
\end{equation}
\begin{equation}
    I_y^{1-\alpha}N_{xy}=0, ~I_y^{1-\alpha}N_{yy}=0, ~I_y^{1-\alpha}Q_{yz}=0, ~I_y^{1-\alpha}M_{xy}=0,
    ~I_y^{1-\alpha}M_{yy}=0 ~~ \forall~\Gamma_y
\end{equation}
\end{subequations}
Note that the natural boundary conditions are nonlocal in nature. This is similar to what is seen in classical integral approaches to the modeling of nonlocal plates \cite{lu2007non}. We anticipate that the nonlocal nature of the natural boundary conditions does not concern us immediately as we will solve the above system of equations using a FE technique. Recall that natural boundary conditions are implicitly satisfied when obtaining the solutions using FE techniques and are accurate up to the order of the specific FEM. Additionally, the following initial conditions are required to obtain the transient response:
\begin{subequations}
\label{eq: Mindlin_IC}
\begin{equation}
    \delta u_0=0, ~\delta v_0=0, ~\delta w_0=0, ~\delta\theta_x=0, ~\delta \theta_y = 0 ~~ \forall~\Omega~\text{at}~t=0
\end{equation}
\begin{equation}
    \delta \dot{u}_0=0, ~\delta \dot{v}_0=0, ~\delta \dot{w}_0=0, ~\delta \dot{\theta}_x=0, ~\delta \dot{\theta}_y = 0 ~~ \forall~\Omega~\text{at}~t=0
\end{equation}
\end{subequations}
Note that the governing equations for the in-plane and transverse displacements are uncoupled, similar to what is seen in the classical Mindlin plate formulation. As expected, the classical plate governing equations and boundary conditions are recovered for $\alpha=1$.
Note that the solution of the above equations yield the mid-plane displacements. The entire displacement field of the plate can then be obtained using Eq.~(\ref{eq: Mindlin_Kinematics}). 

The plate governing equations given in Eq.~(\ref{eq: Mindlin_GDE}) can be expressed in terms of the displacement field variables by using the constitutive stress-strain relations of the plate. For the sake of generality, here below, we provide the expressions for an orthotropic plate:
\begin{subequations}
\label{eq: Mindlin_orthotropic_GDE}
\begin{equation}
A_{11} \mathfrak{D}^\alpha_x\left[D^\alpha_x u_0\right] + 
A_{12} \mathfrak{D}^\alpha_x\left[D^\alpha_y v_0\right] + A_{66} \mathfrak{D}^\alpha_y \left[D^\alpha_y u_0+D^\alpha_x v_0\right] + F_x = I_0\frac{\partial^2 u_0}{\partial t^2}
\end{equation}
\begin{equation}
A_{66} \mathfrak{D}^\alpha_x \left[D^\alpha_y u_0+D^\alpha_x v_0\right] + 
A_{12} \mathfrak{D}^\alpha_y\left[D^\alpha_x u_0\right]  + A_{22} \mathfrak{D}^\alpha_y\left[D^\alpha_y v_0\right] + F_y = I_0\frac{\partial^2 v_0}{\partial t^2}
\end{equation}
\begin{equation}
K_s A_{55}\mathfrak{D}^\alpha_x \left[D^\alpha_x w_0 - \theta_x\right] + K_s A_{44}\mathfrak{D}^\alpha_y \left[D^\alpha_y w_0 - \theta_y\right] + F_z = I_0\frac{\partial^2w_0}{\partial t^2}
\end{equation}
\begin{equation}
D_{11} \mathfrak{D}^\alpha_x\left[D^\alpha_x \theta_x\right] + 
D_{12} \mathfrak{D}^\alpha_x\left[D^\alpha_y \theta_y\right] + D_{66} \mathfrak{D}^\alpha_y \left[D^\alpha_y \theta_x + D^\alpha_x \theta_y\right] + K_s A_{55}\left[D^\alpha_x w_0 - \theta_x\right]
+ I_2\frac{\partial^2\theta_x}{\partial t^2} = M_{\theta_x}
\end{equation}
\begin{equation}
D_{66} \mathfrak{D}^\alpha_x \left[D^\alpha_y \theta_x + D^\alpha_x \theta_y\right] + 
D_{12} \mathfrak{D}^\alpha_y\left[D^\alpha_x \theta_x\right] + D_{22} \mathfrak{D}^\alpha_y\left[D^\alpha_y \theta_y\right] + K_s A_{44}\left[D^\alpha_y w_0 - \theta_y\right]
+ I_2\frac{\partial^2\theta_y}{\partial t^2} = M_{\theta_y}
\end{equation}
\end{subequations}
where the different material constants used in the above equations are given as:
\begin{equation}
    \label{eq: Orthotropic_material_constants_set1}
    A_{11}= \frac{E_1 h}{1-\nu_{12}\nu_{21}}, ~A_{12}=\frac{\nu_{12} E_2 h}{1-\nu_{12}\nu_{21}}, ~A_{22}=\frac{E_2 h}{1-\nu_{12}\nu_{21}}, ~A_{44}=G_{23}h, ~A_{55}=G_{13}h, ~A_{66}=G_{12}h 
\end{equation}
In the above equation, $E_1$ and $E_2$ are the moduli of elasticity along the $\hat{x}$ and $\hat{y}$ axes, respectively. $\nu_{12}$ and $\nu_{21}$ are the Poisson's ratios. Recall that, for an orthotropic material, we have $\nu_{12}/E_1 = \nu_{21}/E_2$. We highlight here that the different subscripts used with the above material constants are consistent with the notations used in literature. Additionally, we also have:
\begin{equation}
    \label{eq: Orthotropic_material_constants_set2}
    D_{11}= \frac{E_1 h^3}{12(1-\nu_{12}\nu_{21})}, ~D_{12}=\frac{\nu_{12} E_2 h^3}{12(1-\nu_{12}\nu_{21})}, ~D_{22}=\frac{E_2 h^3}{12(1-\nu_{12}\nu_{21})}, ~D_{66}=\frac{G_{12}h^3}{12}
\end{equation}
The expressions for the boundary conditions in terms of the displacement variables are given as:
\begin{equation}
\forall~\Gamma_x
\begin{cases}
    \delta u_0=0~~~\text{or}~~~I_x^{1-\alpha}\left[ A_{11}D^\alpha_x u_0 + A_{12} D^\alpha_y v_0\right]=0\\
    \delta v_0=0~~~\text{or}~~~I_x^{1-\alpha}\left[A_{66} \left(D^\alpha_x v_0+D^\alpha_y u_0 \right)\right]=0\\
    \delta w_0=0~~~\text{or}~~~I_x^{1-\alpha}\left[A_{55}\left(D^\alpha_x w_0-\theta_x\right)\right]=0\\
    \delta \theta_x=0~~~\text{or}~~~I_x^{1-\alpha}\left[D_{11} D^\alpha_x \theta_x + D_{12} D^\alpha_y\theta_y\right]=0\\
    \delta \theta_y=0~~~\text{or}~~~I_x^{1-\alpha}\left[D_{66} \left(D^\alpha_y \theta_x+D^\alpha_x\theta_y \right)\right]=0\\
\end{cases}
\end{equation}
\begin{equation}
\forall~\Gamma_y
\begin{cases}
    \delta u_0=0~~~\text{or}~~~I_y^{1-\alpha}\left[A_{66} \left(D^\alpha_x v_0+D^\alpha_y u_0 \right)\right]=0\\
    \delta v_0=0~~~\text{or}~~~I_y^{1-\alpha}\left[D_{12}D^\alpha_x u_0 + D_{22}D^\alpha_y v_0\right]=0\\
    \delta w_0=0~~~\text{or}~~~I_y^{1-\alpha}\left[ A_{44} \left(D^\alpha_y w_0-\theta_y\right)\right]=0\\
    \delta \theta_x=0~~~\text{or}~~~I_y^{1-\alpha}\left[D_{66} \left(D^\alpha_y \theta_x+D^\alpha_x\theta_y \right)\right]=0\\
    \delta \theta_y=0~~~\text{or}~~~I_y^{1-\alpha}\left[D_{12} D^\alpha_x \theta_x + D_{22} D^\alpha_y\theta_y\right]=0\\
\end{cases}
\end{equation}
The governing equations for an isotropic fractional-order plate can be obtained by setting $E_1=E_2=E$, $\nu_{12}=\nu_{21}=\nu$, and $G_{12}=G_{13}=E/2(1+\nu)$ in the above equations.

\section{Fractional-Order Modeling of Nonlocal Kirchoff Plates}
\label{sec: Kirchoff_Plates}
In a similar way to the approach used above, we apply variational principles to develop a fractional-order analogue of the classical Kirchoff plate theory. Recall that the Kirchoff theory is applicable only to thin plates, that is when the plate in-plane characteristic dimension to thickness ratio is on the order of 50 or greater \cite{reddy2006theory}. Under such conditions, the transverse shear strains $\gamma_{xz}$ and $\gamma_{yz}$ can be neglected and the rotations $\theta_x$ and $\theta_y$, introduced previously in \S\ref{sec: Mindlin_Plates}, are approximated as:
\begin{equation}
    \label{eq: Kirchoff_Kinematics}
    \{\theta_x, ~\theta_y \} \approx \left\{\frac{\partial w_0}{\partial x},~\frac{\partial w_0}{\partial y}\right\}
\end{equation}
The strain-displacement relations for the fractional-order Kirchoff plate are derived by substituting the above assumptions in the strain-displacement relations of the fractional-order Mindlin plate given in Eq.~(\ref{eq: Mindlin_Kinematics}). Using the above formalism, the fractional-order strains for the Kirchoff plate are obtained as:
\begin{subequations}
\label{eq: Kirchoff_strains}
\begin{equation}
    \epsilon_{xx}=D^\alpha_x u_0-zD^\alpha_x\left(\frac{\partial w}{\partial x}\right)
\end{equation}
\begin{equation}
    \epsilon_{yy}=D^\alpha_y v_0-zD^\alpha_y\left(\frac{\partial w}{\partial y}\right)
\end{equation}
\begin{equation}
    \gamma_{xy}=D^\alpha_y u_0+D^\alpha_x v_0-z\left[D^\alpha_y \left(\frac{\partial w}{\partial x}\right)+D^\alpha_x \left(\frac{\partial w}{\partial y}\right)\right]
\end{equation}
\end{subequations}
The corresponding stresses are determined using the stress-strain relationships given in Eq.~(\ref{eq: stress_equation}). 
The governing differential equations and the corresponding boundary and initial conditions of the fractional-order Kirchoff plate are derived by using variational principles as illustrated in \S\ref{sec: Mindlin_Plates} for fractional-order Mindlin plates, hence we do not provide the detailed derivation here. It appears that the differential equations governing the linear in-plane and the transverse response are uncoupled similar to the Mindlin plates. Given the nature of the kinematic relations in Eqs.~(\ref{eq: Mindlin_Kinematics},\ref{eq: Kirchoff_Kinematics}), it is immediate that the differential equations and the boundary conditions governing the in-plane response of both the fractional-order Mindlin and the fractional-order Kirchoff plates are identical to each other, similar to what is established in classical plate theories. Hence, we do not provide them explicitly here. However, the governing equation corresponding to the transverse response of the fractional-order Kirchoff plate is different from that of the fractional-order Mindlin plate and it is obtained as:
\begin{equation}
\label{eq: Kirchoff_GDE}
D^{1}_x\left[\mathfrak{D}^\alpha_x M_{xx}\right] + D^1_x \left[\mathfrak{D}^\alpha_y M_{xy} \right] + D^1_y \left[\mathfrak{D}^\alpha_x M_{xy} \right] + D^{1}_y\left[\mathfrak{D}^\alpha_y M_{yy}\right] + F_z = I_0\frac{\partial^2 w_0}{\partial t^2} - I_2 \frac{\partial^2}{\partial t^2}\left[\frac{\partial^2 w_0}{\partial x^2} + \frac{\partial^2 w_0}{\partial y^2}\right]
\end{equation}
The corresponding boundary conditions and initial conditions are obtained as:
\begin{subequations}
\label{eq: Kirchoff_BC}
\begin{equation}
\forall~\Gamma_x
\left\{
\begin{matrix*}[l]
    \delta w_0=0 & \text{or} & \mathfrak{D}^\alpha_x M_{xx} + 2\mathfrak{D}^\alpha_y M_{xy} + I_2 D^1_x \ddot{w}_0 = 0\\
    \delta D^1_x w_0 = 0 & \text{or} & M_{xx}=0
\end{matrix*} \right.
\end{equation}
\begin{equation}
\forall~\Gamma_y
\left\{
\begin{matrix*}[l]
    \delta w_0=0 & \text{or} & \mathfrak{D}^\alpha_y M_{yy} + 2\mathfrak{D}^\alpha_x M_{xy} + I_2 D^1_y \ddot{w}_0 = 0\\
    \delta D^1_y w_0 = 0 & \text{or} & M_{yy}=0
\end{matrix*} \right.
\end{equation}
\begin{equation}
\label{eq: Kirchoff_IC}
    \forall~\Omega~ \text{at}~ t=0,  ~ \delta w_0=0, ~\delta \dot{w}_0=0
\end{equation}
\end{subequations}
In the above equations, $D^{1}_{\square}$ denotes the first-integer order derivative with respect to the spatial variables $\square\in\{x,y\}$, and $\ddot{\square}$ denotes the second integer-order derivative with respect to time.
The governing equations and the corresponding boundary conditions given in Eqs.~(\ref{eq: Kirchoff_GDE},\ref{eq: Kirchoff_BC}) are expressed in terms of the displacement variables for an orthotropic plate as:
\begin{subequations}
\label{eq: Kirchoff_GDE_orthotropic_plate}
\begin{equation}
\begin{split}
D_{11} &\mathfrak{D}^{1+\alpha}_x \left[D^{1+\alpha}_x w_0\right] +
2D_{12} \mathfrak{D}^{1+\alpha}_x \left[ D^{1+\alpha}_y w_0 \right] 
+ D_{22} \mathfrak{D}^{1+\alpha}_y\left[ D^{1+\alpha}_y w_0 \right] +
\\ & D_{66} \left[ D^1_y \mathfrak{D}^\alpha_x ( D^\alpha_x D^1_y w_0 + D^\alpha_y D^1_x w_0) + D^1_x \mathfrak{D}^\alpha_y (D^\alpha_x D^1_y w_0 + D^\alpha_y D^1_x w_0) \right]
=
F_z - I_0\ddot{w}_0 + I_2 \nabla^2 \ddot{w}_0
\end{split}
\end{equation}
\begin{equation}
\forall~\Gamma_x
\left\{
\begin{matrix*}[l]
    \delta w_0=0 & \text{or} & D_{11} \mathfrak{D}^\alpha_x \left[ D^{1+\alpha}_x w_0 \right] + 2D_{66}\left[\mathfrak{D}^\alpha_y \left(D^\alpha_x D^1_y w_0 + D^\alpha_y D^1_x w_0 \right) \right] + \\
     &  &  \hspace{4.6cm} D_{12} \left[ \mathfrak{D}^\alpha_x D^{1+\alpha}_y w_0 \right]  - I_2 D^1_x \ddot{w}_0 = 0\\
    \delta D^1_x w_0 = 0 & \text{or} & D_{11} D^{1+\alpha}_x w_0  + D_{12} D^{1+\alpha}_y w_0 = 0
\end{matrix*} \right.
\end{equation}
\begin{equation}
\forall~\Gamma_y
\left\{
\begin{matrix*}[l]
    \delta w_0=0 & \text{or} & D_{22} \mathfrak{D}^\alpha_y \left[ D^{1+\alpha}_y w_0 \right] + 2D_{66}\left[\mathfrak{D}^\alpha_x \left(D^\alpha_x D^1_y w_0 + D^\alpha_y D^1_x w_0 \right) \right] + \\
     &  &  \hspace{4.6cm} D_{12} \mathfrak{D}^\alpha_y \left[ D^{1+\alpha}_x w_0 \right]  - I_2 D^1_x \ddot{w}_0 = 0\\
    \delta D^1_y w_0 = 0 & \text{or} & D_{12} D^{1+\alpha}_x w_0  + D_{22} D^{1+\alpha}_y w_0 = 0
\end{matrix*} \right.
\end{equation}
\end{subequations}
We highlight here that we have avoided using brackets within the various operators in the above equations for the sake of brevity. We emphasize that the various operators in the above equation are applied sequentially, for example, $D^1_y \mathfrak{D}^\alpha_x D^\alpha_x D^1_y w_0 = D^1_y \left[ \mathfrak{D}^\alpha_x \left\{ D^\alpha_x \left( D^1_y w_0 \right) \right\} \right]$.

\section{2D Fractional Finite Element Method (f-FEM)}
\label{sec: FEM}
The fractional-order nonlocal governing equations for the nonlocal Mindlin and Kirchoff plates are numerically solved via a nonlocal finite element method. The FE formulation developed for the fractional-order governing equations builds upon the FE methods developed in the literature for integral models of nonlocal elasticity \cite{pisano2009nonlocal,phadikar2010variational,norouzzadeh2017finite}. However, several modifications are necessary owing to both the choice and the behavior of the attenuation functions used in the definition of the fractional-order derivatives, as well as to the nonlocal continuum model adopted in this study. In the following, we present the f-FEM for the fractional-order Mindlin theory and then we outline the modifications that would be required in the f-FEM for the fractional-order Kirchoff theory.

\subsection{2D f-FEM Formulation}
\label{sec: ffem_method}
We formulate the f-FEM starting from a discretized form of the Hamiltonian functional given in Eq.~(\ref{eq: extended_hamiltons_principle}).
For this purpose, the plate domain $\Omega=~[0,L]\times[0,B]$ is uniformly discretized into disjoint four-noded quadrilateral (Q4) elements $\Omega^e_i$, such that $\cup_{i=1}^{N_e}\Omega^e_i = \Omega$, and $\Omega^e_j\cap\Omega^e_k=\emptyset~\forall~j\neq k$. $N_e$ is the total number of discretized elements. The Cartesian coordinates $(x,y)$ of each point $\textbf{x} \in\Omega^e_i$ are interpolated by using the $C^0$ Lagrangian shape functions for Q4 elements $\left( \mathcal{L}^{(k)}, k=1,2,3,4 \right)$ in the following manner:
\begin{equation}
    \label{eq: geometric_interpolation}
    (x,y)=\left(\{\mathcal{L}\}_i\{X\}^e_i,\{\mathcal{L}\}_i\{Y\}^e_i\right)
\end{equation}
where $\{\mathcal{L}_i\} = \left\{\mathcal{L}^{(1)}_i ~\mathcal{L}^{(2)}_i ~\mathcal{L}^{(3)}_i ~\mathcal{L}^{(4)}_i \right\}$ is a row vector consisting of the shape functions, while $\{X^e_i\}$ and $\{Y^e_i\}$ are column vectors consisting of the $x$ and $y$ coordinates of the nodes of the element $\Omega^e_i$.
The vector containing the nodal degrees of freedom of the element $\Omega^e_i$ is given as:
\begin{equation}
    \label{eq: Mindlin_DOF}
    \{U^e_i\} = \left\{ \{u_0 ~v_0 ~w_0 ~\theta_x ~\theta_y\}_{i,1} ~\{u_0 ~v_0 ~w_0 ~\theta_x ~\theta_y\}_{i,2} ~\{u_0 ~v_0 ~w_0 ~\theta_x ~\theta_y\}_{i,3} ~\{u_0 ~v_0 ~w_0 ~\theta_x ~\theta_y\}_{i,4} \right\}^T
\end{equation}
where the subscript $(i,k)$ denotes the element number $(i)$ and the local node number $(k)$. The unknown displacement field variables $\{u_0, v_0, w_0, \theta_x, \theta_y\}$ at any point $\textbf{x} \in \Omega^e_i$ are evaluated by interpolating the corresponding nodal degrees of freedom of $\Omega^e_i$. For example, $u_0$ at a point $\textbf{x}\in\Omega^e_i$ can be obtained as:
\begin{equation}
    \label{eq: u_interpolation}
    u_0(\textbf{x}) = \left\{\mathcal{L}_i^{(1)} ~0 ~0 ~0 ~0 ~\mathcal{L}_i^{(2)} ~0 ~0 ~0 ~0 ~\mathcal{L}_i^{(3)} ~0 ~0 ~0 ~0 ~\mathcal{L}_i^{(4)} ~0 ~0 ~0 ~0 \right\} \{U^e_i\} \equiv \left\{ \mathbb{L}^{(u)}_i (\textbf{x}) \right\} \{U^e_i\} 
\end{equation}
The superscript in the row vector $\left\{ \mathbb{L}^{(u)}_i (\textbf{x}) \right\}$ indicates the specific displacement variable being interpolated which is $u_0$ in Eq.~(\ref{eq: u_interpolation}) and the subscript denotes the element number. The other displacement variables can be obtained using similar interpolations. 

We use the discretization scheme discussed above to approximate the Hamiltonian of the system. 
We start by deriving the discretized form of the nonlocal virtual strain energy.
The stress and moment resultants in Eq.~(\ref{eq: Mindlin_virtual_strain_step1}) can be expressed as:
\begin{subequations}
\label{eq: FE_Constitutive_expression}
\begin{equation}
    \resizebox{\textwidth}{!}{$
    \left\{ N_{xx}, N_{yy}, N_{xy}, M_{xx}, M_{yy}, M_{xy} \right\}^T = [S_B]\left\{ D^\alpha_x u_0, D^\alpha_x v_0,  D^\alpha_y u_0 + D^\alpha_x v_0,  D^\alpha_x \theta_x,  D^\alpha_y \theta_y, D^\alpha_y \theta_x + D^\alpha_x \theta_y  \right\}^T$}
\end{equation}
\begin{equation}
    \left\{ Q_{yz}, Q_{xz} \right\}^T = [S_S]\left\{ D^\alpha_y w_0 - \theta_y, D^\alpha_x w_0 - \theta_x  \right\}^T
\end{equation}
\end{subequations}
where $[S_B]$ and $[S_S]$ are the constitutive matrices of the plate. The different elements of these matrices have already been presented for orthotropic materials in Eqs.~(\ref{eq: Orthotropic_material_constants_set1},\ref{eq: Orthotropic_material_constants_set2}). It is immediate that the approximation of the virtual strain energy requires the approximation of the different fractional-order derivatives in Eq.~(\ref{eq: FE_Constitutive_expression}). 

Consider the fractional-order derivative $D^\alpha_x u_0$ at the point $\textbf{x}\in\Omega^e_i$.
Using the definition of the RC derivative given in Eq.~(\ref{eq: RC_definition}), $D_{x}^{\alpha} \left[u_0(\textbf{x})\right]$ is expressed as:
\begin{equation}
    \label{eq: rc_disp_simpl}
    D_{x}^{\alpha} \left[u_0(\textbf{x})\right] = \frac{1}{2}(1-\alpha) \left[l_{A_x}^{\alpha-1} \int_{x-l_{A_x}}^{x} \frac{ D^1_{x^\prime} [u_0(\textbf{x}^\prime)] }{(x-x^\prime)^{\alpha}} ~\mathrm{d}x^\prime + l_{B_x}^{\alpha-1}\int_{x}^{x+l_{B_x}} \frac{ D^1_{x^\prime} [u_0(\textbf{x}^\prime)] }{(x^\prime-x)^{\alpha}}~ \mathrm{d}x^\prime\right]
\end{equation}
where $x^\prime$ is a dummy variable along the $\hat{x}$ axis used within the definition of the fractional-order derivative. The above expression can be recast as:
\begin{subequations}
\label{eq: rc_simpl_form_combined}
\begin{equation}
    \label{eq: rc_simpl_form2}
    D_{x}^{\alpha} \left[u_0(\textbf{x})\right]=\int_{x-l_{A_x}}^{x+l_{B_x}} \mathcal{K}(x, x^\prime, l_{A_x}, l_{B_x}, \alpha)~D^1_{x^\prime} \left[u_0(\textbf{x}^\prime)\right]~\mathrm{d}x^\prime 
\end{equation}
where
\begin{equation}
    \label{eq: frac_kernel_gen}
    \mathcal{K}(x, x^\prime, l_{A_x}, l_{B_x}, \alpha) = \begin{cases}
    \frac{1}{2} (1-\alpha) ~{l_{A_x}}^{\alpha-1} ~{|x - x^\prime|^{-\alpha}} & ~~ x^\prime \in (x-l_{A_x}, x)\\
    \frac{1}{2} (1-\alpha) ~{l_{B_x}}^{\alpha-1} ~{|x - x^\prime|^{-\alpha}} & ~~ x^\prime \in (x, x+ l_{B_x})
\end{cases}
\end{equation}
\end{subequations}
is the kernel of the fractional derivative. 
Note that kernel $\mathcal{K}(x, x^\prime, l_{A_x}, l_{B_x}, \alpha)$ is a function of the relative distance between the points $x$ and $x^\prime$, and can be interpreted similarly to the attenuation functions used in integral models of nonlocal elasticity. Clearly, the attenuation decays as a power-law in the distance with an exponent equal to the order $\alpha$ of the fractional derivative. 
Note that $D^{\alpha}_{x}\left[u_0(\textbf{x})\right]$ contains the integer-order derivative $D^1_{x^\prime}[u_0(\textbf{x}^\prime)]$. $D^1_{x^\prime}[u_0(\textbf{x}^\prime)]$ is evaluated at $\textbf{x}'$ in terms of the nodal displacement variables corresponding to the element $\Omega_p^e$, such that $\textbf{x}'\in \Omega_p^e$. Using Eq.~(\ref{eq: u_interpolation}), the integer-order derivative can be expressed as:
\begin{equation}
\label{eq: deriv_disp_rel}
    D^1_{x'} [u_0(\textbf{x}')] = [B_{u,x}(\textbf{x}')]\{U^e_p\}
\end{equation}
where the $[B_{u,x}(\textbf{x}')]$ is given as:
\begin{equation}
\label{eq: integer_u_b_mat}
    [B_{u,x} (\textbf{x}')] = \left\{\frac{\partial \mathcal{L}_p^{(1)}}{\partial x} ~0 ~0 ~0 ~0 ~\frac{\partial \mathcal{L}_p^{(2)}}{\partial x} ~0 ~0 ~0 ~0 ~\frac{\partial \mathcal{L}_p^{(3)}}{\partial x} ~0 ~0 ~0 ~0 ~\frac{\partial \mathcal{L}_p^{(4)}}{\partial x} ~0 ~0 ~0 ~0 \right\} \equiv \frac{\partial}{\partial x} \left[ \left\{ \mathbb{L}^{(u)}_p (\textbf{x}') \right\} \right]
\end{equation}
The subscript in $[B_{u,x}(\textbf{x}')]$ indicates that the displacement variable under consideration is $u_0$ and the direction of the differentiation is $\hat{x}$.
Using the above expression for the integer-order derivative $D^1_{x'} [u_0(\textbf{x}')]$, the fractional-order derivative in Eq.~\eqref{eq: rc_simpl_form_combined} is obtained as:
\begin{equation}
    \label{eq: rc_disp_simpl_3}
    D_{x}^{\alpha} \left[u_0(\textbf{x})\right] = \int_{x - l_{A_x}}^{x + l_{B_x}}  \mathcal{K}(x, x^\prime, l_{A_x}, l_{B_x}, \alpha) [B_{u,x}(\textbf{x}')] \{U^e_{\textbf{x}'}\} \mathrm{d} x'
\end{equation}
where $\{U^e_{\textbf{x}'}\}$ denotes the vector containing the nodal degrees of freedom of the element $\Omega^e_p$ such that $\textbf{x}' \in \Omega^e_p$.
It is immediate that the evaluation of the fractional derivative in the Eq.~(\ref{eq: rc_disp_simpl_3}) requires a convolution of the integer-order derivative across the interval $(x - l_{A_x}, x + l_{B_x})$. Note that, although the interval of the fractional derivative in Eq.~(\ref{eq: rc_disp_simpl_3}) is $(x - l_{A_x}, x + l_{B_x})$, the horizon of locality at any point $\textbf{x}\in\Omega$ is still two-dimensional in nature. In Eq.~(\ref{eq: rc_disp_simpl_3}), the fractional derivative is being evaluated only in the $\hat{x}$ direction, hence the interval of the derivative is one-dimensional in nature.

While obtaining the FE approximation of the fractional derivative in Eq.~(\ref{eq: rc_disp_simpl_3}), the nonlocal contributions from the different finite elements in the horizon have to be correctly attributed to the corresponding nodes of those elements. In order to correctly account for these nonlocal contributions from the elements in the horizon, we transform the nodal values $\{U^e_{\textbf{x}'}\}$ into the global degrees of freedom vector $\{U\}$ using connectivity matrices in the following manner:
\begin{equation}
\label{eq: conversion_to_global_form}
    \{U^e_{\textbf{x}'}\} = [\tilde{\mathcal{C}}(\textbf{x},\textbf{x}')]\{U\}
\end{equation}
The connectivity matrix $[\tilde{\mathcal{C}}(\textbf{x},\textbf{x}')]$ is designed such that it is non-zero only if the point $\textbf{x}'$ lies in the domain $(\textbf{x}'- \textbf{l}_A , \textbf{x} + \textbf{l}_B)$, that is the horizon of nonlocality for $\textbf{x}$. It is immediate to see that these matrices activate the contribution of the nodes enclosing $\textbf{x}'$ for the numerical evaluation of the convolution integral in Eq. \eqref{eq: rc_disp_simpl_3}. Using the above formalism, Eq.~(\ref{eq: rc_disp_simpl_3}) is expressed as:
\begin{subequations}
\label{eq: frac_der_u_final}
\begin{equation}
    D^{\alpha}_{x}\left[u_0(\textbf{x})\right]=[\tilde{B}_{u,x}(\textbf{x})]\{U\}
\end{equation}
where
\begin{equation}
\label{eq: frac_der_Bu_final}
    [\tilde{B}_{u,x} (\textbf{x})] = \int_{x - l_{A_x}}^{x + l_{B_x}} \mathcal{K}(x, x^\prime, l_{A_x}, l_{B_x}, \alpha) [B_{u,x}(\textbf{x}')] [\tilde{\mathcal{C}}(\textbf{x},\textbf{x}')] ~\mathrm{d}x'
\end{equation}
\end{subequations}
By following the steps summarized by Eqs.~(\ref{eq: rc_disp_simpl}-\ref{eq: frac_der_u_final}), all the remaining fractional derivatives in Eq.~(\ref{eq: FE_Constitutive_expression}) are approximated as:
\begin{subequations}
\label{eq: FE_frac_derivatives_final}
\begin{equation}
    D^{\alpha}_{y}\left[v_0(\textbf{x})\right] = [\tilde{B}_{v,y}(\textbf{x})] \{U\}
\end{equation}
\begin{equation}
    D^{\alpha}_{y}\left[u_0(\textbf{x})\right] +  D^{\alpha}_{x}\left[v_0(\textbf{x})\right] = \left[ [\tilde{B}_{u,y}(\textbf{x})] + [\tilde{B}_{v,x}(\textbf{x})] \right] \{U\}
\end{equation}
\begin{equation}
    D^{\alpha}_{x}\left[\theta_x(\textbf{x})\right] = [\tilde{B}_{\theta_x,x}(\textbf{x})] \{U\}
\end{equation}
\begin{equation}
    D^{\alpha}_{y}\left[\theta_y(\textbf{x})\right] = [\tilde{B}_{\theta_y,y}(\textbf{x})] \{U\}
\end{equation}
\begin{equation}
   D^{\alpha}_{y}\left[\theta_x(\textbf{x})\right] +  D^{\alpha}_{x}\left[ \theta_y (\textbf{x})\right] = \left[ [\tilde{B}_{\theta_x,y}(\textbf{x})] + [\tilde{B}_{\theta_y,x}(\textbf{x})] \right] \{U\}
\end{equation}
\begin{equation}
   D^{\alpha}_{x}\left[w_0(\textbf{x})\right] - \theta_x = \left[ [\tilde{B}_{w,x}(\textbf{x})] - [{\mathbb{L}}^{(\theta_x)}(\textbf{x})] \right] \{U\}
\end{equation}
\begin{equation}
   D^{\alpha}_{y}\left[w_0(\textbf{x})\right] - \theta_y = \left[ [\tilde{B}_{w,y}(\textbf{x})] - [ {\mathbb{L}}^{(\theta_y)}(\textbf{x}) ] \right] \{U\}
\end{equation}
\end{subequations}
$[{\mathbb{L}}^{(\theta_x)}(\textbf{x})]$ and $[{\mathbb{L}}^{(\theta_y)}(\textbf{x})]$ are obtained by assembling the element interpolation vectors for the displacement variables $\theta_x$ and $\theta_y$ given in Eq.~(\ref{eq: u_interpolation}).
By using the above expressions for FE approximation of the different fractional-order derivatives and the constitutive relations given in Eq.~(\ref{eq: FE_Constitutive_expression}), the first variation of the strain energy $\delta\mathcal{U}$ defined in Eq.~\eqref{eq: Mindlin_virtual_strain_step1} is obtained as:
\begin{equation}
\label{eq: FE_expression_virtual_strain}
    \delta \mathcal{U} = \delta \{U\}^T \left[ \int_{\Omega} [\tilde{B}_B(\textbf{x})]^T [S_B] [\tilde{B}_B(\textbf{x})] \mathrm{d}\Omega + \int_{\Omega} [\tilde{B}_S(\textbf{x})]^T [S_S] [\tilde{B}_S(\textbf{x})] \mathrm{d}\Omega \right] \{U\}
\end{equation}
where the matrices $[\tilde{B}_B(\textbf{x})]$ and $[\tilde{B}_S(\textbf{x})]$ are given as:
\begin{subequations}
\label{eq: FE_B_matrices_collection}
\begin{equation}
\begin{split}
        [\tilde{B}_B(\textbf{x})] = \left[ [\tilde{B}_{u,x}(\textbf{x})]^T, [\tilde{B}_{v,y}(\textbf{x})]^T, \left[ [\tilde{B}_{u,y}(\textbf{x})] + [\tilde{B}_{v,x}(\textbf{x})] \right]^T, [\tilde{B}_{\theta_x,x}(\textbf{x})]^T, [\tilde{B}_{\theta_y,y}(\textbf{x})]^T \right., \\ \left. \left[ [\tilde{B}_{\theta_x,y}(\textbf{x})] + [\tilde{B}_{\theta_y,x}(\textbf{x})] \right]^T \right]^T
\end{split}
\end{equation}
\begin{equation}
    [\tilde{B}_S(\textbf{x})] = \left[ \left[ [\tilde{B}_{w,x}(\textbf{x})] - [\tilde{\mathbb{L}}_{\theta_x}(\textbf{x})] \right]^T, \left[ [\tilde{B}_{w,y}(\textbf{x})] - [\tilde{\mathbb{L}}_{\theta_y}(\textbf{x})] \right] \right]^T
\end{equation}
\end{subequations}
By using the interpolations for the displacement fields, the virtual work is approximated as:
\begin{equation}
\label{eq: FE_expression_virtual_work}
    \delta V = \delta \{U\}^T \int_\Omega \left[ \big\{\mathbb{L}^{(u)}\big\}^T F_x + \big\{\mathbb{L}^{(v)}\big\}^T F_y + \big\{\mathbb{L}^{(w)}\big\}^T F_z + \big\{\mathbb{L}^{(\theta_x)}\big\}^T M_{\theta_x} + \big\{\mathbb{L}^{(\theta_y)}\big\}^T M_{\theta_y} \right] \mathrm{d}\Omega
\end{equation}
where the row vectors $\big\{\mathbb{L}^{(\square)}\big\}$ are obtained by assembling the element interpolation vectors given in Eq.~(\ref{eq: u_interpolation}). Similarly, the approximation for the kinetic energy is obtained as:
\begin{equation}
\label{eq: FE_expression_virtual_kinetic_energy}
    \delta T = -\delta \{U\}^T \left[ \int_\Omega \left\{\bar{\mathbb{L}} \right\} \{I_0, ~I_0, ~I_0, ~I_2, ~I_2\}^T \left\{\bar{\mathbb{L}} \right\}^T \mathrm{d}\Omega \right]\{\ddot{U}\}
\end{equation}
where $\left\{\bar{\mathbb{L}} \right\} = \left\{ \big\{\mathbb{L}^{(u)}\big\}^T, \big\{\mathbb{L}^{(v)}\big\}^T,  \big\{\mathbb{L}^{(w)}\big\}^T, \big\{\mathbb{L}^{(\theta_x)}\big\}^T, \big\{\mathbb{L}^{(\theta_y)}\big\}^T \right\}$.
The expressions for $\delta\mathcal{U}, \delta V$ and $\delta K$ given in Eqs.~(\ref{eq: FE_expression_virtual_strain} - \ref{eq: FE_expression_virtual_kinetic_energy}) are substituted in Eq.~(\ref{eq: extended_hamiltons_principle}) and the algebraic equations corresponding to the 2D f-FEM are derived by using the fundamental lemma of variational calculus as:
\begin{equation}
    \label{eq: FE_algebraic_equations}
    [M]\{\ddot{U}\} + [K]\{U\} = \{F\}
\end{equation}
where the mass matrix $[M]$, the stiffness matrix $[K]$, and the force vector $\{F\}$ are given as:
\begin{subequations}
\label{eq: FE_matrices}
\begin{equation}
\label{eq: FE_mass_matrix}
    [M] = \int_\Omega \left\{\bar{\mathbb{L}} \right\} \{I_0, ~I_0, ~I_0, ~I_2, ~I_2\}^T \left\{\bar{\mathbb{L}} \right\}^T \mathrm{d}\Omega
\end{equation}
\begin{equation}
\label{eq: FE_stiffness_matrix}
    [K] = \int_{\Omega} [\tilde{B}_B(\textbf{x})]^T [S_B] [\tilde{B}_B(\textbf{x})] \mathrm{d}\Omega + \int_{\Omega} [\tilde{B}_S(\textbf{x})]^T [S_S] [\tilde{B}_S(\textbf{x})] \mathrm{d}\Omega 
\end{equation}
\begin{equation}
\label{eq: FE_force_vector}
    \{F\} = \int_\Omega \left[ \{\mathbb{L}_u\}^T F_x + \{\mathbb{L}_v\}^T F_y + \{\mathbb{L}_w\}^T F_z + \{\mathbb{L}_{\theta_x}\}^T M_{\theta_x} + \{\mathbb{L}_{\theta_y}\}^T M_{\theta_y} \right] \mathrm{d}\Omega
\end{equation}
\end{subequations}
The solution of the algebraic Eq.~\eqref{eq: FE_algebraic_equations} gives the nodal displacement variables which can then be used along with the kinematic relations in Eq.~(\ref{eq: Mindlin_Kinematics}) to determine the displacement field at any point in the plate.

\subsection{Adaptation to a Fractional-Order Kirchoff Plate}
In the following, we briefly discuss the modifications required in the above FE formulation in order to obtain the response of the fractional-order Kirchoff plate. 

\ul{\textit{Modification \#1}}: it appears from Eq.~(\ref{eq: Kirchoff_Kinematics}) that the FE approximation must ensure continuity of the first-order derivatives of the transverse displacement. Consequently, $C^1$ Hermitian shape functions for Q4 elements have to be used in order to interpolate the transverse displacement field. Recall that Q4 elements are of two types: conforming and non-confirming. In this study, we have used conforming Q4 elements, hence the degrees of freedom corresponding to a node $k$ of the element $\Omega^e_i$ are:
\begin{equation}
    \label{eq: Kirchoff_dof}
    \{U^e_{i,k}\} = \left\{u_0 ~~v_0 ~~w_0 ~~\frac{\partial w_0}{\partial x} ~~\frac{\partial w_0}{\partial y} ~~\frac{\partial^2 w_0}{\partial x \partial y} \right\}_{i,k} 
\end{equation}
The elemental degrees of freedom vector for the element $\Omega^e_i$ can be obtained similar to Eq.~(\ref{eq: Mindlin_DOF}) as $\{U^e_{i}\} = \big\{ \{U^e_{i,1}\}, ~\{U^e_{i,2}\}  \{U^e_{i,3}\}, ~\{U^e_{i,4}\}  \big\}$. The transverse displacement at a point $\textbf{x} \in \Omega^e_i$ can now be interpolated similar to Eq.~(\ref{eq: u_interpolation}) as:
\begin{equation}
    \label{eq: Kirchoff_w_interpolation}
    w_0(\textbf{x}) = \left\{ 0, 0, \mathcal{H}_i^{(1)} \dots \mathcal{H}_i^{(4)}, 0, 0, \mathcal{H}_i^{(5)} \dots \mathcal{H}_i^{(8)}, 0, 0, \mathcal{H}_i^{(9)} \dots \mathcal{H}_i^{(12)}, 0, 0, \mathcal{H}_i^{(13)} \dots \mathcal{H}_i^{(16)} \right\} \{U^e_i\} 
\end{equation}
The interpolating row vector in the above equation is denoted as $\left\{\mathbb{H}^{(w)}_i (\textbf{x}) \right\}$. Note that the in-plane displacement fields are still interpolated using $C^0$ Lagrangian elements.

\ul{\textit{Modification \#2}}: as evident from Eq.~(\ref{eq: Kirchoff_strains}), the evaluation of the discretized strain energy of the fractional-order Kirchoff plate would require the evaluation of the fractional-derivatives of the first integer-order derivatives of the transverse displacement, that is $D^\alpha_{x_2} \left[ D^1_{x_1} w_0 \right]$, where $x_1$ and $x_2$ is either $x$ or $y$. These approximations can be derived by following the steps presented in Eqs.~(\ref{eq: rc_disp_simpl}-\ref{eq: frac_der_u_final}). The FE approximation of $D^\alpha_{x_1} \left[ D^1_{x_2} w_0 \right]$ is given by:
\begin{subequations}
\label{eq: Kirchoff_der_w}
\begin{equation}
    D^{\alpha}_{x_2} \left[ D^1_{x_1} w_0 (\textbf{x}) \right] = [\tilde{B}_{w,{x_1} {x_2}}(\textbf{x})]\{U\}
\end{equation}
\begin{equation}
\label{eq: Kirchoff_der_Bw}
    [\tilde{B}_{w,{x_1} {x_2}}(\textbf{x})] = \int_{x_2 - l_{A_{x_2}}}^{x_2 + l_{B_{x_2}}} \mathcal{K}(x_2, x^\prime_2, l_{A_{x_2}}, l_{B_{x_2}}, \alpha) [B_{w,x_1 x_2}(\textbf{x}')] [\tilde{\mathcal{C}}(\textbf{x},\textbf{x}')] ~\mathrm{d}x^\prime_2
\end{equation}
\begin{equation}
\label{eq: Kirchoff_Bw}
    [{B}_{w,{x_1}{x_2}} (\textbf{x}')] = \frac{\partial^2}{\partial {x'_2} \partial {x'_1}} \left[ \left\{ \mathbb{H}^{(w)}_p (\textbf{x}') \right\} \right]
\end{equation}
\end{subequations}
The subscript $p$ in Eq.~(\ref{eq: Kirchoff_Bw}) is such that the point $\textbf{x}'\in\Omega^e_p$. Similar expressions can be derived for the other fractional-order derivatives of the transverse displacement field. We do not provide them here for the sake of brevity. 

\ul{\textit{Modification \#3}}: the contribution of the transverse shear strains to the stiffness matrix must be removed. More specifically, this can be obtained by setting $[S_S] = 0$ in Eq.~(\ref{eq: FE_stiffness_matrix}) after having carried out the aforementioned modifications. 

\ul{\textit{Modification \#4}}: the expressions for the mass matrix and the force vector must be modified according to the interpolation given in Eq.~(\ref{eq: Kirchoff_w_interpolation}). We do not provide these expressions explicitly as they can be easily found in classical texts discussing FE formulations for plates \cite{reddy2006theory}.

\subsection{Numerical Integration Scheme for the Nonlocal Matrices}
In the following, we provide the details of the numerical scheme used to integrate the stiffness matrix of the fractional-order Mindlin plate given in Eq.~(\ref{eq: FE_stiffness_matrix}). The same procedure directly extends to the evaluation of the stiffness matrix of the fractional-order Kirchoff plate. The procedure to numerically evaluate the mass matrix and the force vector follows directly from classical FE formulations, hence we do not provide a complete description of all the steps. The evaluation of the stiffness matrices for the nonlocal system given in Eq. \eqref{eq: FE_stiffness_matrix} requires the evaluation of the different nonlocal matrices $[\tilde{B}_{\square}], ~ \square \in \{B,S\}$ given in Eq.~(\ref{eq: FE_B_matrices_collection}). As evident from Eqs.~(\ref{eq: frac_der_u_final},\ref{eq: FE_frac_derivatives_final},\ref{eq: FE_B_matrices_collection}), this involves a convolution of the integer-order derivatives with the fractional-order attenuation function over the horizon of nonlocality. Clearly, the FE approximation for fractional-order derivatives involves additional integrations over the horizon of nonlocality to account for the nonlocal interactions.
A numerical procedure to account for these nonlocal interactions was presented in \cite{polizzotto2001nonlocal,pisano2009nonlocal}. Differently from these studies, the attenuation function in the fractional-order model involves an end-point singularity due to the nature of the power-law kernel (see Eq.~(\ref{eq: frac_kernel_gen})). The fractional-order nonlocal interactions as well as the end-point singularity are addressed in \cite{patnaik2019FEM} where a fractional-order FEM has been developed for modeling 1D nonlocal beams. Here below, we briefly review this numerical procedure.

In the following, we describe the procedure to numerical evaluate the contribution due to the bending stress and moment resultants in the stiffness matrix $[K]$ given in Eq.~(\ref{eq: FE_stiffness_matrix}) and denoted as $[K_B]$ in the following. The same procedure directly extends to evaluate the contributions of the transverse shear resultants. We adopt an isoparametric formulation and introduce a natural coordinate system $(\xi,\eta)$ to numerically integrate $[K_B]$. The Jacobian of the transformation $(x,y) \rightarrow (\xi,\eta)$ is given as $J(\xi,\eta)$. By using the Gauss-Legendre quadrature rule, the matrix $[K_B]$ is approximated as:
\begin{equation}
    [K_B]
    \approx \sum\limits_{i=1}^{N_e}\sum\limits_{j=1}^{N_{GP}} \hat{w}_j J^i \left[\tilde{B}_B \left(\xi^{i,j}, \eta^{i,j} \right) \right]^T [S_B] \left[\tilde{B}_B \left(\xi^{i,j}, \eta^{i,j} \right) \right]^T
\end{equation}
where $(\xi^{i,j}, \eta^{i,j})$ is the $j-$th Gauss-Legendre point in the $i-$th element, $\hat{w}_j$ is the corresponding weight for numerical integration, and $N_{GP}$ is the total number of Gauss points chosen for the numerical integration, such that $j\in \{1,...N_{GP}\}$. $J^i$ is the Jacobian of the coordinate transformation for the $i-$th element. The hat symbol on the weight is used to distinguish it from the transverse displacement.
As previously highlighted, the matrices $[\tilde{B}_{\square}]$ ($\square\in\{B,S\}$) involve a convolution integration due to the fractional-order nonlocality. Note from Eq.~(\ref{eq: FE_B_matrices_collection}) that $[\tilde{B}_{\square}]$ contains the different nonlocal matrices given in Eq.~(\ref{eq: FE_frac_derivatives_final}). In the following we outline the procedure for the evaluation of the matrix $[\tilde{B}_{u,x}(\textbf{x})]$ only. The same procedure extends directly for the evaluation of the remaining matrices in Eq.~(\ref{eq: FE_B_matrices_collection}). $[\tilde{B}_{u,x}(\textbf{x})]$ is approximated as:
\begin{equation}
\label{eq: B_numerical_step0}
    \left[\tilde{B}_{u,x}(\textbf{x}^{i,j})\right] \equiv \left[\tilde{B}_{u,x}(x^{i,j},y^{i,j})\right] = \int_{x^{i,j} - l_{A_x}}^{x^{i,j} + l_{B_x}} \mathcal{K}(x, x^\prime, {l}_{A}, {l}_{B}, \alpha) [B_{u,x}(\textbf{x}')] [\tilde{\mathcal{C}_\square}(\textbf{x}^{i,j}, \textbf{x}')] ~\mathrm{d}x'
\end{equation}
where $(x^{i,j},y^{i,j})$ is the Cartesian coordinate of the Gauss point $(\xi^{i,j}, \eta^{i,j})$ and $[B_{\square}(s_1)]$ is given in Eq.~\eqref{eq: integer_u_b_mat}. Using Eq.~\eqref{eq: frac_kernel_gen}, we obtain the following expression for $\left[\tilde{B}_{u,x}(\textbf{x}^{i,j})\right]$:
\begin{subequations}
\label{eq: B_numerical_step1}
\begin{equation}
\label{eq: b_mat_numer_simp1}
    \left[\tilde{B}_{u,x}(x^{i,j},y^{i,j})\right] = \int_{x^{i,j}-l_{A_x}}^{x^{i,j}} \mathcal{I}_L \mathrm{d}x' + \int_{x^{i,j}}^{x^{i,j}+l_{B_x}} \mathcal{I}_R \mathrm{d}x'
\end{equation}
where, the integrands $\mathcal{I}_L$ and $\mathcal{I}_R$ are obtained by substituting the expression for the fractional derivative kernel given in Eq.~(\ref{eq: frac_kernel_gen}) in the above Eq.~(\ref{eq: B_numerical_step0}) as:
\begin{equation}
    \mathcal{I}_L = \frac{1}{2} (1-\alpha) ~{l_{A_x}}^{\alpha-1} ~{|x - x^\prime|^{-\alpha}} [B_{u,x}(\textbf{x}')] [\tilde{\mathcal{C}}(\textbf{x}^{i,j}, \textbf{x}')]
\end{equation}
\begin{equation}
    \mathcal{I}_R =\frac{1}{2} (1-\alpha) ~{l_{B_x}}^{\alpha-1} ~{|x - x^\prime|^{-\alpha}} [B_{u,x}(\textbf{x}')] [\tilde{\mathcal{C}}(\textbf{x}^{i,j}, \textbf{x}')]
\end{equation}
\end{subequations}
Note that the terminals of the integrals in Eq.~\eqref{eq: b_mat_numer_simp1} span over the elements that constitute the nonlocal horizon along the $\hat{x}$ direction: $(x^{i,j} - l_{A_x}, x^{i,j} + l_{B_x})$ at $\textbf{x}^{i,j}$. These integrals are evaluated numerically in the following manner:
\begin{subequations}
\label{eq: b_mat_int_Scheme}
\begin{equation}
    \int_{x^{i,j}-l_{A_x}}^{x^{i,j}} \mathcal{I}_L \mathrm{d}x' \approx \underbrace{\int_{x^{i-N_{A_x}^{inf}}}^{x^{i - N_{A_x}^{inf} + 1}} \mathcal{I}_L \mathrm{d}x' + ...\int_{x^{i-1}}^{x^{i}} \mathcal{I}_L \mathrm{d}x' }_{\text{Gauss-Legendre Quadrature}}+ \underbrace{\int_{x^{i}}^{x^{i,j}} \mathcal{I}_L \mathrm{d}x' }_{\text{Singularity at } x^{i,j}}
\end{equation}
\begin{equation}
    \int_{x^{i,j}}^{x^{i,j}+l_{B_x}} \mathcal{I}_R \mathrm{d}x' \approx \underbrace{\int_{x^{i,j}}^{x^{i+1}} \mathcal{I}_R \mathrm{d}x' }_{\text{Singularity at } x^{i,j}} + \underbrace{\int_{x^{i+1}}^{x^{i+2}} \mathcal{I}_R \mathrm{d}x' ... + \int_{x^{i + N_{B_x}^{inf}-1}}^{x^{i+N_{B_x}^{inf}}} \mathcal{I}_R \mathrm{d}x' }_{\text{Gauss-Legendre Quadrature}}
\end{equation}
\end{subequations}
In the above expressions, $N_{A_x}^{inf}$ and $N_{B_x}^{inf}$ are the number of (complete) elements in the nonlocal horizon to the left and right side of the point $\textbf{x}$ along the $\hat{x}$ direction, respectively. More specifically, $N_{A_x}^{inf}=\ceil{l_{A_x}/l_{e_x}}$ and ${N_{B_x}^{inf}}=\lfloor{l_{B_x}/l_{e_x}}\rfloor$ where $l_{e_x}$ is the dimension of the discretized element along the $\hat{x}$ (assuming a uniform discretization along the $\hat{x}$ direction). The ceil ($\ceil{\cdot}$) and floor ($\lfloor{\cdot}\rfloor$) functions are used to round the number of elements to the greater integer on the left side and the lower integer on the right side. For points $\textbf{x}^{i,j}$ close to the boundaries $\Gamma_x$ of the plate ($x=\{0,L\}$), $N_{A_x}^{inf}$ and $N_{B_x}^{inf}$ are truncated in order to account for asymmetric horizon lengths. This is essential to satisfy frame-invariance of the formulation as discussed in \S\ref{sec: Nonlocal Elasticity via Fractional Calculus}.

As discussed previously, due to the nature of the kernel of the fractional-order derivative, an end-point singularity occurs in the integrals at the Gauss point $\textbf{x}^{i,j}$ in the element $\Omega_i^e$. This is evident from the definitions of the left and right integrals given in Eq.~(\ref{eq: B_numerical_step1}). Following \cite{patnaik2019FEM}, this end-point singularity is circumvented by an analytical evaluation of these integrals over the elements containing the singularities. This analytical evaluation can be carried out by using the expression for $[\tilde{B}_{u,x}]$ given in Eq.~(\ref{eq: frac_der_Bu_final}).
The integrals over the remaining elements (i.e. those without singularities) are evaluated using the Gauss-Legendre quadrature method. The expression for this integration corresponding to the nonlocal contribution of the $r-$th element in the horizon of nonlocality of the Gauss point $\textbf{x}^{i,j}$ along the $\hat{x}$ direction is given as:
\begin{equation}
\label{nonlocal_integral_example}
\begin{split}
        \int_{x^{r}}^{x^{r+1}} \mathcal{K}(x, x^\prime, {l}_{A}, {l}_{B}, \alpha) [B_{u,x}(\textbf{x}')] & [\tilde{\mathcal{C}_\square}(\textbf{x}^{i,j}, \textbf{x}')] ~\mathrm{d}x' = \\
    & \sum_{k=1}^{N_{GP}}\hat{w}_k J^r \mathcal{K}(x, x^\prime, {l}_{A}, {l}_{B}, \alpha) [B_{u,x}({\textbf{x}}^{r,k})] [\tilde{\mathcal{C}}(\textbf{x}^{i,j}, {\textbf{x}}^{r,k})]
\end{split}
\end{equation}
where $x^{r,k}$ is the Cartesian coordinate of the $k-$th Gauss point in the $r-$th element along $\hat{x}$, $\hat{w}_k$ is the corresponding weight, and $J^r$ is the Jacobian of the transformation for $r-$th element. We highlight that the attenuation kernel $\mathcal{K}$ is a function of the Cartesian coordinates, and therefore global coordinates should be used in its evaluation. 

\section{Validation and Convergence of the 2D f-FEM}
\label{sec: Validation_convergence}
In this section, we present the results of a validation and convergence study carried out for the 2D f-FEM approach. For this purpose, we fixed the in-plane dimensions of the plate to be $L=1$m and $B=1$m. The thickness of the plate was taken to be $0.1$m $(=L/10)$ for the fractional-order Mindlin formulation, and $0.01$m $(=L/100)$ for the fractional-order Kirchoff formulation. The material was assumed isotropic with an elastic modulus $E=30$GPa for both cases, while a Poisson's ratio $\nu=0.3$ was chosen for the fractional-order Mindlin plate and $\nu=0.25$ for the fractional-order Kirchoff plate. We emphasize that although the numerical results were obtained for an isotropic plate, the f-FEM formulation developed in \S\ref{sec: FEM} is applicable to any type of medium through appropriate changes in the constitutive matrices $[S_B]$ and $[S_S]$ defined in Eq.~(\ref{eq: FE_Constitutive_expression}). In the following, we have assumed a symmetric and isotropic horizon of nonlocality for points sufficiently inside the domain of the plate, that is $l_{A_\square} = l_{B_\square} = l_f, \square\in\{x,y\}$. For points located close to a boundary, the length scales are truncated appropriately as discussed in \S\ref{sec: Nonlocal Elasticity via Fractional Calculus}.

\subsection{Validation of FEM Model}
\label{ssec: validation}
\begin{figure*}[ht!]
    \centering
    \begin{subfigure}[t]{0.49\textwidth}
        \centering
        \includegraphics[width=\textwidth]{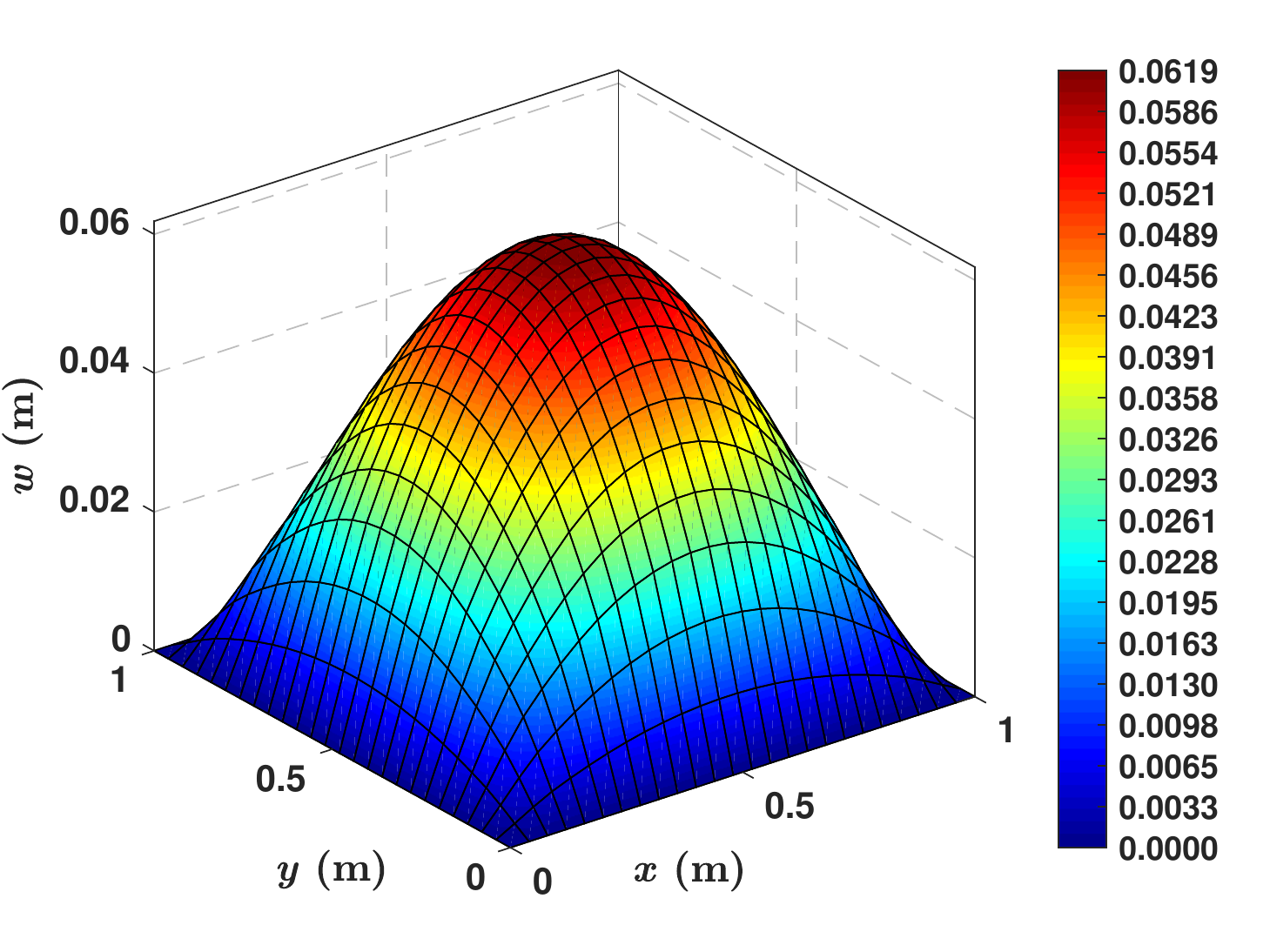}
        \caption{Numerical simulation obtained via 2D f-FEM}
    \end{subfigure}%
    ~ 
    \begin{subfigure}[t]{0.49\textwidth}
        \centering
        \includegraphics[width=\textwidth]{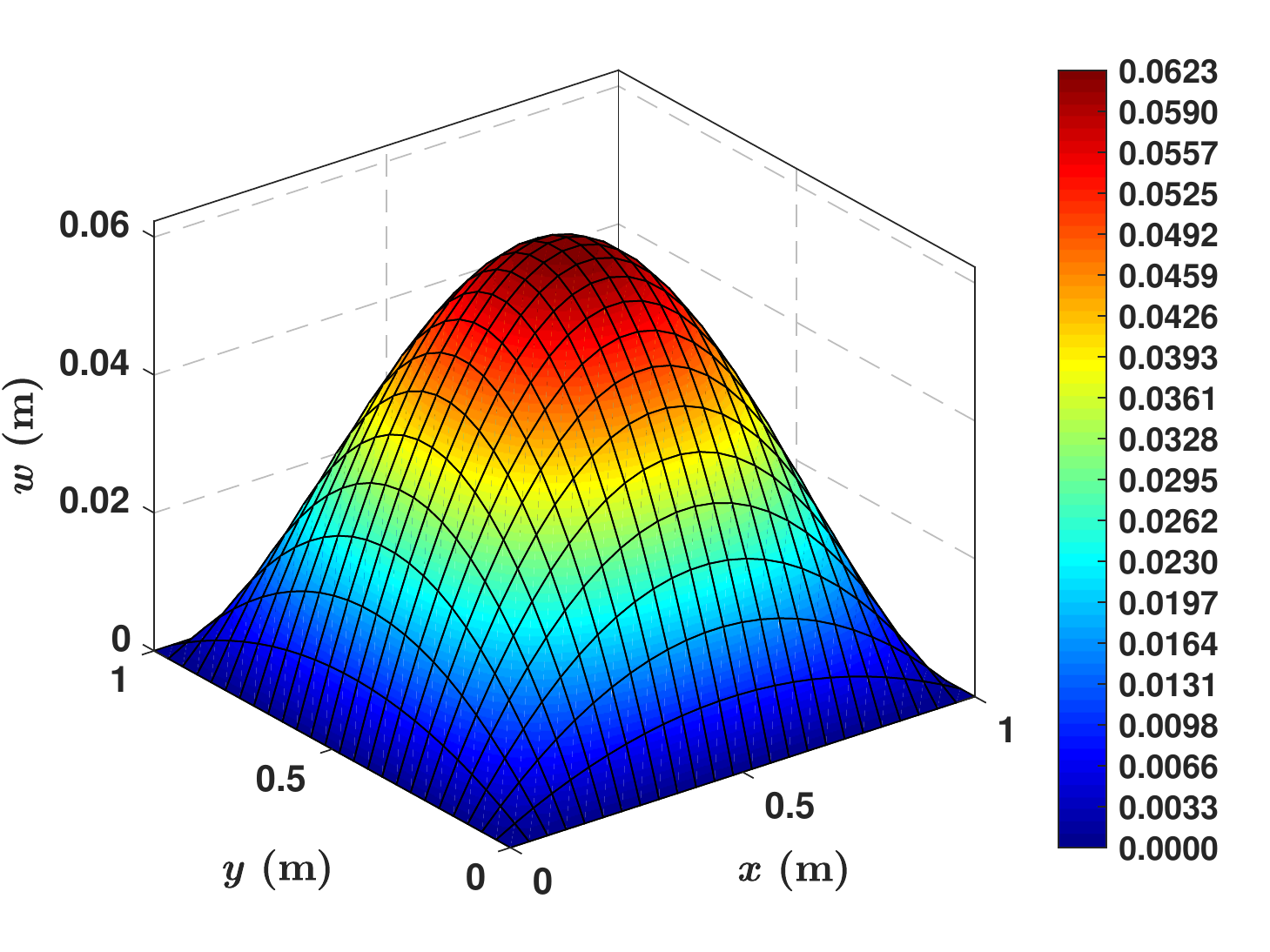}
        \caption{Exact Solution}
    \end{subfigure}
    \caption{Comparison of the transverse displacement of the fractional-order Mindlin plate with $\alpha=0.8$ and $l_f=0.1L$ clamped at all its edges obtained via (a) the 2D f-FEM and (b) the exact solution given in Eq.~(\ref{eq: validation_assumption}). As evident from the color maps the match between the numerically obtained solution and the exact solution is excellent.}
    \label{fig: validation_plot}
\end{figure*}

We validate the 2D f-FEM by comparing the results of the f-FEM against the exact solution of a fractional-order plate clamped all around its boundaries and subject to spatially varying loads. More specifically, the following displacement field of the mid-plate of the plate is assumed:
\begin{equation}
\label{eq: validation_assumption}
    u_0 = 0 ~~v_0 = 0 ~~w_0 = xy(x - 1)(y - 1) ~~\theta_x = xy(x - 1)(y - 1) ~~\theta_y = xy(x - 1)(y - 1)
\end{equation}
Note that the above displacement field satisfies the boundary conditions for a plate clamped at its edges. The strong form of the governing differential equations in Eq.~(\ref{eq: Mindlin_GDE}) is used to obtain the loads required to satisfy the displacement response in Eq.~(\ref{eq: validation_assumption}). We highlight here that the assumed displacement field is independent of the order $(\alpha)$ and of the length scale $(l_f)$, hence resulting in a forcing function which is dependent on the fractional parameters. The obtained force field is then used within the f-FEM and the corresponding numerical approximation of the transverse displacement provided by the f-FEM is compared against the exact solution given in Eq.~(\ref{eq: validation_assumption}). Different combinations of $\alpha$ and $l_f$ were considered. 
The numerical results in terms of the maximum transverse displacement of the plate which is obtained at $(L/2,B/2)$ are given in Table~[\ref{tab: 2D_validation}]. Additionally, we have also provided the plots of the numerically obtained transverse displacement and the exact solution for the specific case of $\alpha=0.8$ and $l_f=0.1L$ in Fig.~(\ref{fig: validation_plot}). As evident from Table [\ref{tab: 2D_validation}], the match is excellent and the error is less than $4\%$ in all the cases. The study corresponding to $\alpha=1$ can be considered as a check on the robustness of the f-FEM.

\begin{table}[h!]
    \centering
        \caption{Numerical results of the maximum transverse displacement predicted by f-FEM results. The results are compared with the maximum transverse displacement obtained by the exact solution. Results are compared for different order $\alpha$ and size of the horizon $l_f$. Note from Eq.~(\ref{eq: validation_assumption}) that we have assumed a displacement which is independent of $\alpha$ and $l_f$. To simplify the comparison of the data, the maximum transverse displacement was scaled by a factor of 100.}
    \label{tab: 2D_validation}
    \begin{tabular}{c|c|c c|c}
    \hline\hline
    \multirow{2}{2em}{~~$\alpha$} & \multirow{2}{2em}{~~$l_f$} &\multicolumn{2}{c|}{{$100\times w~\text{(in m)}$}} & \multirow{2}{3em}{~Error \\ {~~$(\%)$}}\\
    \cline{3-4}
    &  & f-FEM &  Exact & \\
    \hline
       1  & - & 6.26 & 6.25 & 0.16 \\
       0.9  & $0.1L$ & 6.06 & 6.25 & 3.04 \\
       0.9  & $0.2L$ & 6.50 & 6.25 & 4.00 \\
       0.8  & $0.1L$ & 6.19 & 6.25 & 0.96 \\
       0.8  & $0.2L$ & 6.26 & 6.25 & 0.16 \\
       \hline\hline
    \end{tabular}
\end{table}

\subsection{Convergence}
\label{ssec: convergence}
This section presents the results of the sensitivity analysis on the FE mesh size. The convergence of the integer-order FEM with element discretization, referred to as the $h-$refinement, is well established in the literature. In this study we noted that, in addition to the FE mesh size, the convergence of the f-FEM also depends on the fractional-order and on the ratio of the FE mesh size and the nonlocal length scale. It appears that the convergence of the f-FEM depends on the strength of the nonlocal interactions across the nonlocal horizon. Therefore, sufficient number of elements $\mathcal{N}_x \times \mathcal{N}_y \equiv (l_f/l_{e_x} \times l_f/l_{e_y})$ should be available in the nonlocal horizon to accurately represent the fractional-order nonlocal interactions. $l_{e_x}$ and $l_{e_y}$ denote the size of the FE mesh in the $\hat{x}$ and $\hat{y}$ directions, respectively. Additionally, mesh refinement across the domain of the plate would increase the spatial resolution, hence decreasing inconsistencies due to the truncation of the nonlocal horizon. This approach would provide a better numerical approximation for the nonlocal matrices $[\tilde{B}_\square]$ in Eqs.~(\ref{eq: B_numerical_step1},\ref{eq: b_mat_int_Scheme}). Therefore, convergence of the f-FEM is expected when the number of elements in the nonlocal horizon $\mathcal{N}_x \times \mathcal{N}_y$, referred to as the “dynamic rate of convergence” \cite{norouzzadeh2017finite,patnaik2019FEM}, is sufficient to accurately capture the fractional-order nonlocal interactions. 

In the context of the current study, we establish the convergence of the f-FEM for both the fractional-order Mindlin and Kirchoff plates. The fractional-order plates are clamped on their edges (denoted as CCCC) and under the effect of a uniformly distributed transverse load (UDTL) $F_z$. Results are presented in Table~[\ref{tab: convergence_mindlin}] and Table~[\ref{tab: convergence_Kirchoff}] for different combinations of $\alpha$ and $l_f$. More specifically, the maximum transverse displacement $(w_0(L/2,B/2))$ is compared (moving from top to bottom within a column) for a given fractional-order $(\alpha)$ and length scale $(l_f)$. The maximum displacement in the Tables~[\ref{tab: convergence_mindlin},\ref{tab: convergence_Kirchoff}] is non-dimensionalized in the following manner:
\begin{equation}
    \label{eq: normalized_transverse_disp}
    \overline{w} = w_0 \left(L/2,B/2\right) \left[\frac{100 E h^3}{F_z L^4}\right]
\end{equation}

\begin{table}[h!]
    \centering
    \begin{tabular}{c | c |c c c c}
    \hline\hline
       \multirow{2}{6em}{~~~~~~~$l_f$} & \multirow{2}{6em}{~~~$\mathcal{N}_{x} \times \mathcal{N}_{y}$} & \multicolumn{4}{c}{$\overline{w}$}\\
         \cline{3-6}
         &  & $\alpha=1.0$& $\alpha=0.9$& $\alpha=0.8$ & $\alpha=0.7$ \\
         \hline\hline
         \multirow{4}{5em}{$l_f=0.2L$} & $4\times4$ & 1.5645 & 1.7214 & 1.8549 & 1.9793 \\
         & $8\times8$ & 1.6226 & 1.8164 & 1.9924 & 2.1668 \\
         & $10\times10$ & 1.6299 & 1.8350 & 2.0226 & 2.2092 \\
         & $12\times12$ & 1.6339 & 1.8480 & 2.0445 & 2.2400 \\
         & $16\times16$ & 1.6379 & 1.8659 & 2.0753 & 2.2832 \\
         \hline
         \multirow{4}{5em}{$l_f=0.4L$} & $4\times4$ & 1.3715 & 1.8071 & 2.3811 & 3.1851 \\
         & $8\times8$ & 1.5645 & 2.0525 & 2.6554 & 3.3890 \\
         & $10\times10$ & 1.5836 & 2.0788 & 2.6856 & 3.4102 \\
         & $12\times12$ & 1.6071 & 2.1118 & 2.7252 & 3.4427 \\
         & $16\times16$ & 1.6226 & 2.1384 & 2.7611 & 3.4776 \\
         \hline
         \multirow{4}{5em}{$l_f=0.5L$} & $4\times4$ & 1.2578 & 1.7969 & 2.6313 & 4.0621 \\
         & $8\times8$ & 1.5238 & 2.1893 & 3.2054 & 4.9160 \\
         & $10\times10$ & 1.5645 & 2.2486 & 3.2874 & 5.0246 \\
         & $12\times12$ & 1.5876 & 2.2835 & 3.3362 & 5.0887 \\
         & $16\times16$ & 1.6113 & 2.3227 & 3.3935 & 5.1657 \\
          \hline\hline
    \end{tabular}
    \caption{Convergence of the f-FEM for a CCCC-Mindlin plate for various fractional parameters.}
    \label{tab: convergence_mindlin}
\end{table}

\begin{table}[h!]
    \centering
    \begin{tabular}{c | c |c c c c}
    \hline\hline
       \multirow{2}{6em}{~~~~~~~$l_f$} & \multirow{2}{6em}{~~~$\mathcal{N}_{x} \times \mathcal{N}_{y}$} & \multicolumn{4}{c}{$\overline{w}$}\\
         \cline{3-6}
         &  & $\alpha=1.0$& $\alpha=0.9$& $\alpha=0.8$ & $\alpha=0.7$ \\
         \hline\hline
         \multirow{4}{5em}{$l_f=0.2L$} & $4\times4$ & 1.4235 & 1.5102 & 1.5860 & 1.6603 \\
         & $8\times8$ & 1.4235 & 1.5129 & 1.5916 & 1.6689 \\
         & $10\times10$ & 1.4235 & 1.5135 & 1.5929 & 1.6709 \\
         & $12\times12$ & 1.4235 & 1.5140 & 1.5939 & 1.6273 \\
         \hline
         \multirow{4}{5em}{$l_f=0.4L$} & $4\times4$ & 1.4235 & 1.7262 & 2.0782 & 2.5063  \\
         & $8\times8$ & 1.4235 & 1.7102 & 2.0320 & 2.4078 \\
         & $10\times10$ & 1.4235 & 1.6982 & 2.0135 & 2.3790 \\
         & $12\times12$ & 1.4235 & 1.7055 & 2.0185 & 2.3788 \\
         \hline
         \multirow{4}{5em}{$l_f=0.5L$} & $4\times4$ & 1.4236 & 1.8189 & 2.3280 & 3.0251 \\
         & $8\times8$ & 1.4325 & 1.8072 & 2.2952 & 2.9598 \\
         & $10\times10$ & 1.4235 & 1.8047 & 2.2878 & 2.9446 \\
         & $12\times12$ & 1.4235 & 1.8030 & 2.2828 & 2.9340 \\
         \hline\hline
    \end{tabular}
    \caption{Convergence of the f-FEM for a CCCC-Kirchoff plate for various fractional parameters.}
    \label{tab: convergence_Kirchoff}
\end{table}

It was found that, targeting an error threshold less than 2\% between successive refinements, the dynamic rate of convergence is $\mathcal{N}_x \times \mathcal{N}_y = 12 \times 12$ for the fractional-order Mindlin plate and $\mathcal{N}_x \times \mathcal{N}_y = 10 \times 10$ for the fractional-order Kirchoff plate. We highlight here that the same condition on the dynamic rate of convergence was also observed for plates simply supported on all edges (denoted as SSSS) and subject to a UDTL. These results were not provided here for the sake of brevity. Since the convergence in the displacements is less than 2\% for the aforementioned mesh of $\mathcal{N}_x \times \mathcal{N}_y$, in the following we use this mesh discretization in order to carry out the static and dynamic FE simulations.

\section{Static Response of the Fractional-Order Plates}
\label{sec: static_response}
Having established the accuracy and consistency of the 2D f-FEM, the f-FEM was used to analyze the static response of the fractional-order nonlocal plates. More specifically, we analyzed the effect of the fractional model parameters $\alpha$ and $l_f$ on the response of the fractional-order plates. The dimensions and material properties of the plates were the same used in \S\ref{sec: Validation_convergence}.
Consider a fractional-order Mindlin plate subject to a UDTL. The response of the plate was analyzed for two different kinds of boundary conditions (CCCC and SSSS) and for different combinations of the fractional model parameters. The results of this study are summarized in Table~[\ref{tab: static_response_mindlin_CC}] and Table~[\ref{tab: static_response_mindlin_SS}] in terms of the non-dimensionalized maximum transverse displacement which is obtained at the center of the plate. We analyzed also the static response of a fractional-order Kirchoff plate subjected to a UDTL under the same two boundary conditions (CCCC and SSSS). The results are presented in Table~[\ref{tab: static_response_kirchoff_CC}] and Table~[\ref{tab: static_response_kirchoff_SS}]. The results corresponding to the classical integer order plate models (i.e. $\alpha=1$) presented in Tables~[\ref{tab: static_response_kirchoff_CC},\ref{tab: static_response_kirchoff_SS}] match well with the results given in \cite{reddy2006theory}. The displacement obtained in each case is non-dimensionalized as given in Eq.~(\ref{eq: normalized_transverse_disp}).

\begin{table}[h]
    \centering
    \begin{tabular}{c|c c c c}
    \hline\hline
       \multirow{2}{6em}{~~~~~~~$l_f$} & \multicolumn{4}{c}{$\overline{w}$} \\
         \cline{2-5}
         & $\alpha=1.0$ & $\alpha=0.9$& $\alpha=0.8$ & $\alpha=0.7$ \\
         \hline\hline
         {$l_f=0.2L$} & 1.6071 & 1.8480 & 2.0445 & 2.2400 \\
         {$l_f=0.3L$} & 1.6071 & 1.9554 & 2.2787 & 2.5372 \\
         {$l_f=0.4L$} & 1.6071 & 2.1118 & 2.7252 & 3.4427 \\
         {$l_f=0.5L$} & 1.6071 & 2.2835 & 3.3362 & 5.0887 \\
         \hline\hline
    \end{tabular}
    \caption{Comparison of the effect of the fractional model parameters on the static response of a CCCC-Mindlin plate subjected to a UDTL. Results are presented in terms of the non-dimensionalized maximum transverse displacement.}
    \label{tab: static_response_mindlin_CC}
\end{table}

\begin{table}[h]
    \centering
    \begin{tabular}{c|c c c c}
    \hline\hline
       \multirow{2}{6em}{~~~~~~~$l_f$} & \multicolumn{4}{c}{$\overline{w}$} \\
         \cline{2-5}
         & $\alpha=1.0$ & $\alpha=0.9$& $\alpha=0.8$ & $\alpha=0.7$ \\
         \hline\hline
         {$l_f=0.2L$} & 4.6401 & 5.1533 & 5.5759 & 5.9310 \\
         {$l_f=0.3L$} & 4.6401 & 5.3579 & 6.0026 & 6.5009 \\
         {$l_f=0.4L$} & 4.6401 & 5.6047 & 6.6445 & 7.6796 \\
         {$l_f=0.5L$} & 4.6401 & 5.9198 & 7.6192 & 9.9868 \\
         \hline\hline
    \end{tabular}
    \caption{Comparison of the effect of the fractional model parameters on the static response (in terms of the non-dimensionalized maximum displacement) of a SSSS-Mindlin plate subjected to a UDTL.}
    \label{tab: static_response_mindlin_SS}
\end{table}

\begin{table}[h]
    \centering
    \begin{tabular}{c|c c c c}
    \hline\hline
       \multirow{2}{6em}{~~~~~~~$l_f$} & \multicolumn{4}{c}{$\overline{w}$} \\
         \cline{2-5}
         & $\alpha=1.0$ & $\alpha=0.9$& $\alpha=0.8$ & $\alpha=0.7$ \\
         \hline\hline
         {$l_f=0.2L$} & 1.4235 & 1.5135 & 1.5929 & 1.6709 \\
         {$l_f=0.3L$} & 1.4235 & 1.6047 & 1.7772 & 1.9380 \\
         {$l_f=0.4L$} & 1.4235 & 1.6982 & 2.0135 & 2.3790 \\
         {$l_f=0.5L$} & 1.4235 & 1.8047 & 2.2878 & 2.9446 \\
         \hline\hline
    \end{tabular}
    \caption{Comparison of the effect of the fractional model parameters on the static response (in terms of the non-dimensionalized maximum displacement) of a CCCC-Kirchoff plate subjected to a UDTL.}
    \label{tab: static_response_kirchoff_CC}
\end{table}

\begin{table}[h!]
    \centering
    \begin{tabular}{c|c c c c}
    \hline\hline
       \multirow{2}{6em}{~~~~~~~$l_f$} & \multicolumn{4}{c}{$\overline{w}$} \\
         \cline{2-5}
         & $\alpha=1.0$ & $\alpha=0.9$& $\alpha=0.8$ & $\alpha=0.7$ \\
         \hline\hline
         {$l_f=0.2L$} & 4.5701 & 4.6151 & 4.6419 & 4.6610 \\
         {$l_f=0.3L$} & 4.5701 & 4.7068 & 4.8252 & 4.8768 \\
         {$l_f=0.4L$} & 4.5701 & 4.8249 & 5.0927 & 5.3828 \\
         {$l_f=0.5L$} & 4.5701 & 4.9480 & 5.4094 & 6.0180 \\
         \hline\hline
    \end{tabular}
    \caption{Comparison of the effect of the fractional model parameters on the static response (in terms of the non-dimensionalized maximum displacement) of a SSSS-Kirchoff plate subjected to a UDTL.}
    \label{tab: static_response_kirchoff_SS}
\end{table}

Recall that the classical integral approaches to nonlocal elasticity lead to paradoxical predictions either of hardening or of absence of nonlocal interactions for certain combinations of boundary conditions \cite{khodabakhshi2015unified}. As established in \cite{romano2017constitutive,romano2018nonlocal,barretta2019stress}, the paradoxical predictions in the classical integral approaches result from the mathematically ill-posed nature of the integral constitutive relations. More specifically, the constitutive relation between the bending field and the curvature is a Fredholm integral of the first kind whose solution does not generally exist and, if it exists, it is not necessarily unique \cite{romano2017constitutive,romano2018nonlocal}. In \cite{patnaik2019FEM} it was established that the fractional-order nonlocal model gives rise to a self-adjoint and positive-definite system accepting a unique solution.
Further, as evident from the Tables~[\ref{tab: static_response_mindlin_CC}-\ref{tab: static_response_kirchoff_SS}], the maximum transverse displacement of the plates increases with a decreasing value of the fractional-order $\alpha$ as well as with an increasing value of the length scale $l_f$. It follows that the stiffness of the fractional-order plates decreases due to the increasing degree of nonlocality achieved via either reducing $\alpha$ or increasing $l_f$. Note that this stiffness reduction occurs regardless of the nature of the boundary conditions. 

\section{Free Vibration Response of Fractional-Order Plates}
\label{sec: eigen_response}
This section presents the effect of nonlocality on the natural frequency of vibration of fractional-order plates. The natural frequencies are obtained by solving the eigenvalue problem:
\begin{equation}
    \label{eq: eigen_system}
    [M]^{-1}[K]\{\mathbb{U}\}=\omega_0^2\{\mathbb{U}\}
\end{equation}
which was obtained by assuming a periodic solution $\{U\}=\{\mathbb{U}\}e^{-i\omega_0t}$ to the homogeneous part of the algebraic FE Eq.~(\ref{eq: FE_algebraic_equations}). In the above assumed solution, $i=\sqrt{-1}$, $\omega_0$ denotes the natural frequency of vibration, and $\mathbb{U}$ is the amplitude of the harmonic oscillation. Similar to \S\ref{sec: static_response}, we have obtained the natural frequencies of both the fractional-order Mindlin and Kirchoff plates for different combinations of the fractional model parameters $\alpha$ and $l_f$. Also in this case, two types of boundary conditions (CCCC and SSSS) were considered for both plates. Results are presented in Tables~[\ref{tab: eigen_mindlin_CC}-\ref{tab: eigen_kirchoff_SS}] in terms of the fundamental frequency of transverse vibration. The fundamental frequencies have been non-dimensionalized in the following manner:
\begin{subequations}
\label{eq: normalized_frequency}
\begin{equation}
    \overline{\omega}_0 = \omega_0 \left[ \frac{L^2}{h} \sqrt{\frac{\rho}{E}} \right]~~ \text{(for CCCC)}
\end{equation}
\begin{equation}
    \overline{\omega}_0 = \omega_0 \left[ \left(\frac{B}{\pi}\right)^2 \sqrt{\frac{\rho h}{D_{11}}} \right]~~ \text{(for SSSS)}
\end{equation}
\end{subequations}
where $\rho$ is the density of the plate. 

As evident from Tables~[\ref{tab: eigen_mindlin_CC}-\ref{tab: eigen_kirchoff_SS}], the introduction of the fractional-order nonlocality leads to a decrease in the fundamental frequency of vibration. This result is a direct consequence of the fact that the effective stiffness of the structure decreases due to the nonlocality, as established in \S\ref{sec: static_response}. More specifically, the fundamental frequency of vibration decreases upon increasing the degree of nonlocality, that is either by reducing $\alpha$ or by increasing $l_f$. This reduction in the fundamental frequency due to the nonlocality is consistent with studies conducted in literature by using classical approaches to nonlocal elasticity \cite{hosseini2013exact,pradhan2009nonlocal}. Note that, when using the fractional order formulation, the decrease in the fundamental frequency occurs regardless of the nature of the boundary conditions. 

\begin{table}[h!]
    \centering
    \begin{tabular}{c|c c c c}
    \hline\hline
       \multirow{2}{6em}{~~~~~~~$l_f$} & \multicolumn{4}{c}{$\overline{\omega}_0$} \\
         \cline{2-5}
         & $\alpha=1.0$ & $\alpha=0.9$& $\alpha=0.8$ & $\alpha=0.7$ \\
         \hline\hline
         {$l_f=0.2L$} & 9.8540 & 9.2083 & 8.6610 & 8.1801 \\
         {$l_f=0.3L$} & 9.8540 & 8.9162 & 8.0603 & 7.2857 \\
         {$l_f=0.4L$} & 9.8540 & 8.6172 & 7.4342 & 6.3439 \\
         {$l_f=0.5L$} & 9.8540 & 8.3622 & 6.8856 & 5.5204 \\
         \hline\hline
    \end{tabular}
    \caption{Comparison of the effect of the fractional model parameters on the fundamental frequency of the transverse vibration of a Mindlin plate clamped on all edges.}
    \label{tab: eigen_mindlin_CC}
\end{table}

\begin{table}[h!]
    \centering
    \begin{tabular}{c|c c c c}
    \hline\hline
       \multirow{2}{6em}{~~~~~~~$l_f$} & \multicolumn{4}{c}{$\overline{\omega}_0$} \\
         \cline{2-5}
         & $\alpha=1.0$ & $\alpha=0.9$& $\alpha=0.8$ & $\alpha=0.7$ \\
         \hline\hline
         {$l_f=0.2L$} & 5.7788 & 5.4588 & 5.2210 & 5.0368 \\
         {$l_f=0.3L$} & 5.7788 & 5.3664 & 5.0286 & 4.7443 \\
         {$l_f=0.4L$} & 5.7788 & 5.2581 & 4.7980 & 4.3866 \\
         {$l_f=0.5L$} & 5.7788 & 5.1487 & 4.5570 & 3.9966 \\
         \hline\hline
    \end{tabular}
    \caption{Comparison of the effect of the fractional model parameters on the fundamental frequency of the transverse vibration of a Mindlin plate simply supported on all edges.}
    \label{tab: eigen_Mindlin_SS}
\end{table}

\begin{table}[h!]
    \centering
    \begin{tabular}{c|c c c c}
    \hline\hline
       \multirow{2}{6em}{~~~~~~~$l_f$} & \multicolumn{4}{c}{$\overline{\omega}_0$} \\
         \cline{2-5}
         & $\alpha=1.0$ & $\alpha=0.9$& $\alpha=0.8$ & $\alpha=0.7$ \\
         \hline\hline
         {$l_f=0.2L$} & 3.6457 & 3.5176 & 3.4090 & 3.3077 \\
         {$l_f=0.3L$} & 3.6457 & 3.4123 & 3.2115 & 3.0277 \\
         {$l_f=0.4L$} & 3.6457 & 3.3157 & 3.0288 & 2.7659 \\
         {$l_f=0.5L$} & 3.6457 & 3.2389 & 2.8812 & 2.5483 \\
         \hline\hline
    \end{tabular}
    \caption{Comparison of the effect of the fractional model parameters on the fundamental frequency of the transverse vibration of a Kirchoff plate clamped on all edges.}
    \label{tab: eigen_kirchoff_CC}
\end{table}

\begin{table}[h!]
    \centering
    \begin{tabular}{c|c c c c}
    \hline\hline
       \multirow{2}{6em}{~~~~~~~$l_f$} & \multicolumn{4}{c}{$\overline{\omega}_0$} \\
         \cline{2-5}
         & $\alpha=1.0$ & $\alpha=0.9$& $\alpha=0.8$ & $\alpha=0.7$ \\
         \hline\hline
         {$l_f=0.2L$} & 1.9998 & 1.9865 & 1.9765 & 1.9676 \\
         {$l_f=0.3L$} & 1.9998 & 1.9681 & 1.9384 & 1.9081 \\
         {$l_f=0.4L$} & 1.9998 & 1.9465 & 1.8923 & 1.8342 \\
         {$l_f=0.5L$} & 1.9998 & 1.9257 & 1.8463 & 1.7564 \\
         \hline\hline
    \end{tabular}
    \caption{Comparison of the effect of the fractional model parameters on the fundamental frequency of the transverse vibration of a Kirchoff plate simply supported on all edges.}
    \label{tab: eigen_kirchoff_SS}
\end{table}

We also analyzed the effect of the fractional-order nonlocality as well as of the fractional model parameters on the higher order frequencies. We considered two specific cases: (\#1) we fixed the length-scale at $l_f=0.5L$ and obtained the first eight natural frequencies of transverse vibration for a CCCC-Mindlin plate for different values of $\alpha$; and (\#2) we fixed the fractional-order at $\alpha=0.8$ and obtained the first eight natural frequencies of transverse vibration of a CCCC-Mindlin plate for different values of $l_f$. The results corresponding to the cases \#1 and \#2 are presented in Table [\ref{tab: nonlocal_effect_on_higher_modes_fixed_L}] and Table [\ref{tab: nonlocal_effect_on_higher_modes_fixed_order}], respectively. In both cases, the frequency values corresponding to the classical local plate (i.e. $\alpha=1$) are also presented to provide a reference to estimate the effect of the fractional-order nonlocality. As evident from the Tables [\ref{tab: nonlocal_effect_on_higher_modes_fixed_L},\ref{tab: nonlocal_effect_on_higher_modes_fixed_order}], the fractional-order nonlocality has a stronger effect on the higher vibration modes when compared to the fundamental mode. This can be attributed to the more complex spatial distribution of the strain field produced by the shorter wavelengths associated with higher frequency modes. We emphasize that similar trends were also observed for simply supported Mindlin plates as well as Kirchoff plates. This observation is consistent with the results from classical approaches to the modeling of transverse vibration of nonlocal plates \cite{wang2011mechanisms}.

\begin{table}[h!]
    \centering
    \begin{tabular}{c|c c c c c c c c}
    \hline\hline
         {$\alpha$} & $\overline{\omega}_0$ & $\overline{\omega}_1$ & $\overline{\omega}_2$ & $\overline{\omega}_3$ & $\overline{\omega}_4$ & $\overline{\omega}_5$ & $\overline{\omega}_6$ & $\overline{\omega}_7$ \\
         \hline\hline
         {$\alpha=0.7$} & 5.5204 & 7.8808 & 10.1654 & 11.9102 & 12.0238 & 13.9883 & 17.2615 & 17.3539  \\
         {$\alpha=0.8$} & 6.8856 & 10.9440 & 14.7220 & 16.8717 & 17.0305 & 20.2794 & 24.7794 & 25.2272 \\
         {$\alpha=0.9$} & 8.3622 & 14.6749 & 20.2387 & 23.4535 & 23.6739 & 28.3423 & 34.6228 & 35.5065  \\
         \hline\hline
         {$\alpha=1$} & 9.8540 & 18.8806 & 26.4485 & 31.3387 & 31.6403 & 37.8018 & 46.4625 & 47.7089 \\
         \hline\hline
    \end{tabular}
    \caption{Comparison of the effect of the fractional-order on the first eight frequencies of vibration of a Mindlin plate clamped on all edges. The length scale is fixed at $l_f=0.5L$.}
    \label{tab: nonlocal_effect_on_higher_modes_fixed_L}
\end{table}

\begin{table}[h!]
    \centering
    \begin{tabular}{c|c c c c c c c c}
    \hline\hline
         {$l_f$} & $\overline{\omega}_0$ & $\overline{\omega}_1$ & $\overline{\omega}_2$ & $\overline{\omega}_3$ & $\overline{\omega}_4$ & $\overline{\omega}_5$ & $\overline{\omega}_6$ & $\overline{\omega}_7$ \\
         \hline\hline
         {$l_f=0.4L$} & 7.4342 & 11.7452 & 15.8992 & 16.8007 & 16.9583 & 20.6869 & 24.3174 & 25.0889 \\
         {$l_f=0.3L$} & 8.0603 & 13.3021 & 18.3123 & 18.4040 & 18.5473 & 23.2114 & 24.2828 & 27.8966 \\
         {$l_f=0.2L$} & 8.6610 & 15.2968 & 21.3305 & 22.4971 & 22.6767 & 28.1549 & 29.2432 & 34.3695 \\
         \hline\hline
         {$\alpha=1$} & 9.8540 & 18.8806 & 26.4485 & 31.3387 & 31.6403 & 37.8018 & 46.4625 & 47.7089 \\
         \hline\hline
    \end{tabular}
    \caption{Comparison of the effect of the length scale on the first eight frequencies of vibration of a Mindlin plate clamped on all edges. The fractional-order is fixed at $\alpha=0.8$.}
    \label{tab: nonlocal_effect_on_higher_modes_fixed_order}
\end{table}

\section{Conclusions}
\label{sec: Conclusions}
This paper presented a fractional-order nonlocal plate theory based on a frame-invariant and dimensionally consistent fractional-order nonlocal continuum theory. Nonlocality accounted for by using fractional-order kinematic relations. This approach lead to an explicit relationship between the nonlocal stresses and strains thereby enabling the application of variational principles in order to derive the strong form of the governing equations. The proposed approach also resulted in a self-adjoint and positive definite system which guarantees a unique solution. These latter properties allowed the reformulation of the governing equations in finite element form, which is very convenient to achieve numerical solutions. The fractional order finite element model (f-FEM) was developed by using the Hamilton's principle, hence following an energy minimization strategy which, in this case, was applied on a global scale due to the nonlocal nature of the system. This approach resulted in  pre-assembled global system matrices. 
The proposed f-FEM approach was validated with benchmark problems drawn from both fractional-order equations and nonlocal elasticity. Then, the model was applied to study both the static and the free vibration dynamic response of fractional-order nonlocal plates, in either Mindlin or Kirchoff formulation. It was observed that the nonlocality results in a softening of the structure leading to larger transverse displacements and lower frequency of vibration. The results were shown to be independent of the nature of both the loading and the boundary conditions, hence relieving some typical inconsistencies emerging in classical nonlocal theories when dealing combinations of boundary conditions. 
In conclusion, the results presented in this study illustrated several unique features of fractional calculus for the modeling of nonlocal structures and suggested that this mathematical tool could play a critical role in the development of advanced simulation techniques for complex systems.

\section{Acknowledgements}
The authors gratefully acknowledge the financial support of the Defense Advanced Research Project Agency (DARPA) under grant \#D19AP00052, and of the National Science Foundation (NSF) under grants MOMS \#1761423 and DCSD \#1825837. The content and information presented in this manuscript do not necessarily reflect the position or the policy of the government. The material is approved for public release; distribution is unlimited.

\bibliographystyle{unsrt}
\bibliography{report}
\end{document}